# Single-Atom Catalysis: Insights from Model Systems

Florian Kraushofer and Gareth S. Parkinson*

Institute of Applied Physics, TU Wien, 1040 Vienna, Austria

**Abstract**

The field of single atom catalysis (SAC) has expanded greatly in recent years. While there has been much success developing new synthesis methods, a fundamental disconnect exists between most experiments and the theoretical computations used to model them. The real catalysts are based on powder supports, which inevitably contain a multitude of different facets, different surface sites, defects, hydroxyl groups, and other contaminants due to the environment. This makes it extremely difficult to determine the structure of the active SAC site using current techniques. To be tractable, computations aimed at modelling SAC utilize periodic boundary conditions and low index facets of an idealized support. Thus the reaction barriers and mechanisms determined computationally represent, at best, a plausibility argument, and there is a strong chance that some critical aspect is omitted. One way to better understand what is plausible is by experimental modelling, i.e. comparing the results of computations to experiments based on precisely defined single-crystalline supports prepared in an ultrahigh vacuum (UHV) environment. In this article, we review the status of the surface science literature as it pertains to SAC. We focus on experimental work on supports where the site of the metal atom are unambiguously determined from experiment, in particular the surfaces of rutile and anatase $TiO_2$, the iron oxides $Fe_2O_3$ and $Fe_3O_4$, as well as $CeO_2$ and $MgO$. Much of this work is based on scanning probe microscopy in conjunction with spectroscopy, and we highlight the remarkably few studies in which metal atoms are stable on low index surfaces of typical supports. In a perspective, we discuss the possibility for expanding such studies into other relevant supports such as N-doped carbon, graphene and metal carbides.

parkinson@iap.tuwien.ac.at

List of Acronyms

| | |
|---|---|
| AFM | Atomic Force Microscopy |
| DFT | Density Functional Theory |
| $E_F$ | Fermi level |
| EPR | Electron Paramagnetic Resonance |
| EXAFS | Extended X-ray Absorption Fine Structure |
| HAADF | High-Angle Annular Dark-Field [STEM] |
| IRAS | Infrared Reflection Absorption Spectroscopy |
| KPFM | Kelvin Probe Force Microscopy |
| LEED | Low Energy Electron Diffraction |
| LEED-$I(V)$ | Quantitative Low Energy Electron Diffraction ($I$ intensity, $V$ acceleration voltage) |
| ML | Monolayer |
| MvK | Mars-van Krevelen [mechanism] |
| ncAFM | Non-contact Atomic Force Microscopy |
| PES | Photoelectron Spectroscopy |
| PROX | Preferential Oxidation of CO |
| UHV | Ultrahigh Vacuum |
| SAC | Single-Atom Catalysis |
| SCV | Subsurface Cation Vacancy |
| SRPES | Synchrotron Radiation Photoelectron Spectroscopy |
| STEM | Scanning Transmission Electron Microscopy |
| STM | Scanning Tunnelling Microscopy |
| SXRD | Surface X-ray diffraction |
| [S]TEM | [Scanning] Transmission Electron Microscopy |
| TPD | Temperature Programmed Desorption |
| $V_O$ | Oxygen Vacancy |

XANES         X-ray Absorption Near Edge Structure

XPS             X-ray Photoelectron Spectroscopy

Table of Contents Figure

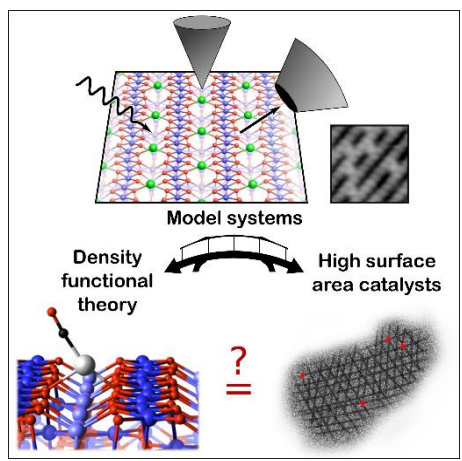

**Authors Bios**

Gareth Parkinson received his PhD in physics at the University of Warwick, UK. He worked as a postdoctoral researcher at PNNL and Tulane University in the USA before moving to TU Wien in Vienna as an assistant professor in 2010. Since 2021 he is full professor of surface reactivity in the Institute of Applied Physics, TU Wien.

Florian Kraushofer received his PhD in physics from the Technical University of Vienna in 2021 under the supervision of Gareth Parkinson, where he worked on iron oxide surfaces as model systems for single atom catalysis. He is now a postdoctoral researcher at the Technical University of Munich.

**Table of Contents**

# 1. Introduction

The field of "single-atom" catalysis has expanded rapidly over recent years with highly efficient and active catalysts demonstrated for a wide variety of chemical,[1-6] photochemical,[7,8] and electrochemical[9] reactions. While the concept seems well established by the sheer number of studies and has been extensively reviewed,[4,10-24] in many cases it is not really clear if, and how, the single atom really catalyzes the reaction.[11] Recent advances in transmission electron microscopy have made it possible to routinely demonstrate the existence of isolated heavy atoms on an as-synthesized catalyst,[25] and the location of the heavy atom can be determined relative to the lattice of the support. One must remember, however, that the TEM image is a 2D projection of a 3D object, so it is not possible to know whether any atom is located at the surface without multiple projections. Moreover, the support lattice show columns of atoms from the bulk of the material, not the surface atoms to which the metal atom is bound. For the oft-used metal oxide supports, the oxygen sublattice is typically not resolved at all, and it is to these atoms that the metal atom are proposed to bind. Even if they were visible, metal-oxide surfaces are often not simple truncations of the bulk structure (i.e. they reconstruct to a minimum energy configuration), and typically contain a variety of different defects. Such sites are thought to bind metal adatoms strongly (knowledge derived from surface-science type studies), and thus likely play a significant role in SAC. Further guidance on the type of site comes from complementary spectroscopies such as XANES, IRAS, and XPS, but these area averaging techniques do not necessarily give information on the active site (which may be a minority species), and are somewhat indirect, as will be discussed in this review.

A second major issue in SAC research is that even if the state of the as-synthesized catalyst can be determined, proving that the system remains atomically dispersed during catalytic reactions remains challenging.[11,26,27] It is possible that the system evolves in the reactive environment to form small nanoparticles, and that these are really the active site. "Post mortem" imaging of samples is not routinely performed, but even in cases where it is, doubts linger as to whether the system could have redispersed once outside the reaction environment. As a consequence, SAC remains controversial and there remains significant scope for fundamental insights.

Ultimately, atomic-scale details regarding the active sites and reaction mechanisms are proposed primarily on the basis of density functional theory (DFT) calculations. Periodic slab calculations based on a low index facet of the support material are used, which may or may not appear on the powder catalyst. A suitable adsorption site for the metal adatom is then commonly selected based on a strong binding energy relative to other possible sites in DFT, with some guidance from experiment. For example, if CO-IRAS measurements suggest the metal is cationic, and XANES suggests coordination to oxygen, then cation-like sites on or within an idealized surface may be tested. With a site selected, the reaction pathway is studied, and a mechanism is proposed. Given the assumptions made about the nature of the support surface and the educated guess at an adsorption site, these calculations represent a plausibility argument, which shows that the reaction could proceed in this way. It is not proof that it does so. Similar caveats exist for the results of theoretical screening studies, which attempt to determine the best metal atom for a particular reaction, because the result depends strongly on the site and mechanism assumed, as seen recently for CO oxidation on $FeO_x$-supported SACs.[28-30]

Clearly then, the complexity of the catalyst makes it difficult to assess whether the site and reaction pathways proposed on the basis of theory are realistic. In this review, we will cover the pertinent literature from the surface-science community that can help to understand how SACs work. In the surface science approach, a single-crystalline sample exhibiting a low-index surface orientation is prepared under UHV conditions (typically by $Ar^+$ sputtering and high temperature annealing) until it is free of contamination such as OH groups and carbon. In this state, the system resembles an experimental analogue of that created by periodic slab calculations. Crucially, the atomic-scale structure of the surface has been determined with sub-angstrom precision for several common support materials such as $TiO_2$, $CeO_2$, MgO, $Fe_3O_4$ and $Fe_2O_3$. While such "model" surfaces necessarily lack some of the complexity of applied SAC systems, they can serve as a solid basis for experimental studies of SAC mechanisms, and as a benchmark for the theoretical approach used in high surface area catalytic studies. Typically, the metal atoms are evaporated directly onto the surface, meaning there are no ligands, and no calcination or activation of the system performed prior to study.

In what follows we will discuss what work exists in the surface science literature that is relevant to SAC, ordered by different support materials. Five main sections give an overview of $TiO_2$, the iron oxides $FeO_x$, $CeO_2$, $MgO$ and $Cu_2O$. At the end of each section, we summarize the state of the surface science research for SACs on a given oxide, including key takeaways. The section on $TiO_2$ is further split into rutile and anatase $TiO_2$, while the section on iron oxides in turn addresses three different $FeO_x$ facets α-$Fe_2O_3$(0001), α-$Fe_2O_3$(0001) and $Fe_3O_4$(001). When discussing adatoms on the oxides that have been most extensively studied by surface science methods ($TiO_2$, $Fe_3O_4$ and $CeO_2$), we loosely group elements by position in the periodic table, or by similar observed behavior of adatoms. Finally, in a Perspective section, we summarize the state of research on these different model systems, discuss it in the context of other work, and give an overview of promising directions for further research.

# 2. Titania (TiO$_2$)

## 2.1     Rutile TiO$_2$(110)

Rutile TiO$_2$(110) is one of the most intensively studied systems in surface science.[31-33] This is partly because single crystal samples of high quality are inexpensive and widely available, and partly because UHV preparation can be easily achieved by in-situ cycles of inert gas sputtering followed by annealing. This treatment results in slight reduction of the sample, which is sufficient to allow experiments based on electron transfer (STM, XPS etc.). The structure of the as-prepared surface is precisely known from quantitative structural techniques (SXRD/LEED-*I*(*V*)),[34,35] and the results agree with DFT calculations[36] and fit well with the results of scanning probe studies[37]. Interestingly, the contrast observed in STM is reversed from the topography, and the low lying Ti$_{5c}$ atoms are imaged bright and the protruding "bridging" O$_{2c}$ rows are imaged dark (see Figure 1a,b).[31] This is related to the electronic structure, and the fact that the Ti atoms have electronic states close to the Fermi level (E$_F$) while the oxygen atoms do not. Specifically, the samples are n-type due to the sample reduction, and the Ti states are located at the conduction band minimum. The newest generation of scanning probe instruments allows simultaneous imaging by STM and non-contact AFM, allowing to image both the atomic and electronic structure of the surface. Due to the reduction of the sample, surface oxygen vacancies (V$_O$) are common defects. They are imaged as bright protrusions in the dark O$_{2c}$ rows in STM, and as missing atoms in the O$_{2c}$ rows in ncAFM (Figure 1a and 1c, respectively). The reduced surface is often denoted r-TiO$_2$(110). One of the most important conclusions from the work on TiO$_2$(110) has been the importance of V$_O$ as active sites for both chemical reactions and the nucleation of metal nanoparticles.[31-33] For example, water molecules react with the oxygen vacancies creating two "surface hydroxyl" groups, i.e. a pair of H atoms adsorbed at the O$_{2c}$ rows. This reaction is highly efficient, and the residual water in a UHV environment is sufficient to fill all the V$_O$ sites over several hours. When saturation exposure to water creates a partially hydroxylated surface, it is often referred to as h-TiO$_2$(110). Similarly, exposing the r-TiO$_2$(110) surface to very small amounts of molecular oxygen results in the repair of the vacancy and the adsorption of an O atom atop the Ti$_{5c}$ rows. The resulting surface is then termed o-TiO$_2$(110). It is important to note that realistic catalytic environments will thus not have surface V$_O$ present in an appreciable number.

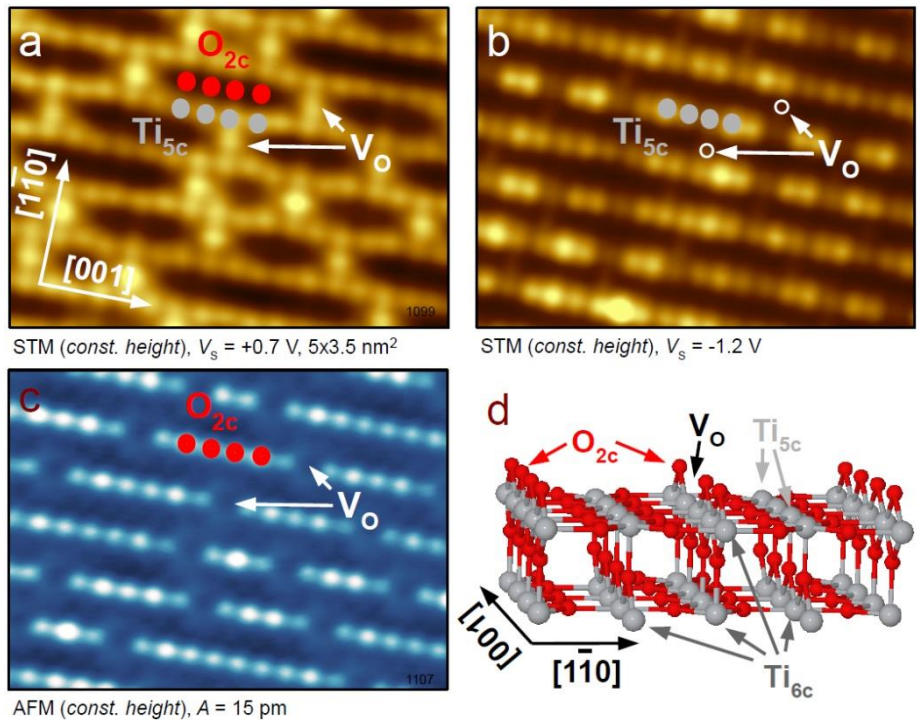

**Figure 1:** STM and nc-AFM imaging of the rutile TiO$_2$ (110) surface. The same area of the sample is shown with (a) empty-states STM, (b) filled states STM, and (c) AFM. All images were measured at a sample temperature T=78 K. (d) A structural model of the surface. The STM and AFM images were measured in constant height mode. Figure adapted from Ref. 37. Copyright 2017 American Physical Society under CC-BY license (https://creativecommons.org/licenses/by/4.0/).

### 2.1.1    Cu, Ag, Au on TiO$_2$(110)

It seems well established that Au$_1$ prefers to bind at V$_O$ sites on r-TiO$_2$(110).[38,39] Thornton and coworkers[39] imaged the V$_o$ sites directly by STM (see Figure 2), and then observed Au atoms to occupy exactly these positions after deposition in UHV. Moreover, they performed voltage pulses with the STM tip which caused the Au atom to hop out of the vacancy, which could be clearly observed after the Au had left.

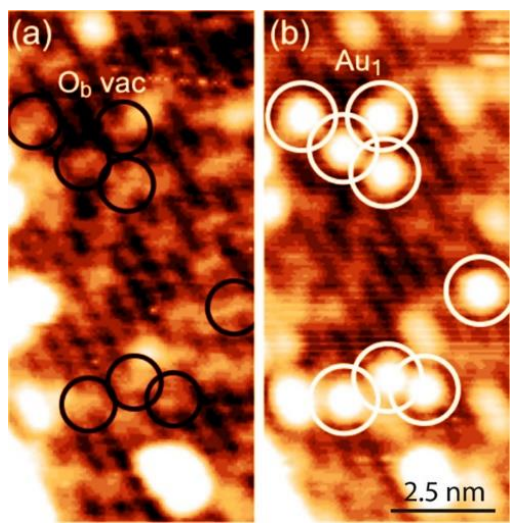

**Figure 2:** STM images of the as-prepared $TiO_2(110)$ surface with oxygen vacancies [black circles in (a)] and following deposition of Au metal at room temperature. The Au atoms clearly occupy the original $V_O$ sites. Figure reproduced with permission from Ref. 39. Copyright 2017 American Chemical Society under CC-BY license (https://creativecommons.org/licenses/by/4.0/).

Recently, a very nice experiment/theory paper revisited the $Au_1/TiO_2(110)$ system with a particular focus on the photocatalytic properties.[40] The authors confirmed that $Au_1$ adatoms preferentially occupy $V_O$ sites, and once these were filled, observed that $Ti_{5c}$ sites became occupied at 80 K. Interestingly, the authors mention that the adsorption configuration obtained in DFT is affected by the size of the computational supercell, as well as by the inclusion of an $V_O$ in the model. With a $V_O$ concentration corresponding to the experimental conditions, a vertical Ti-Au bond is obtained rather than the tilted geometry obtained previously on stoichiometric slabs.[41] A particularly interesting aspect of this study was the measurement of localized metal-induced gap states below $E_F$ by scanning tunneling spectroscopy, which the authors show provides a dedicated channel for the transfer of a photoexcited hole from the $TiO_2$ substrate to the adsorbed Au atoms. The hole transfer could be accomplished by UV light exposure or by the STM tip, and was found to weaken the Ti–Au bond at the 5-fold coordinated Ti site, allowing the Au atoms to diffuse across the surface at 80 K. Atoms adsorbed at the $V_O$ sites were unaffected by UV exposure.

Other interesting insights into adatom behavior can be inferred from surface-science studies of nanoparticle nucleation. A comparison of Au [42] and Ag [43] nanoparticle growth on the *r*-, *h*-, and *o*-TiO$_2$(110) surfaces was performed by Wendt and coworkers, who found that the presence of hydroxyl groups accelerates sintering for both metals. Interestingly, a surface oxygen adatom on the o-TiO$_2$(110) surface was suggested to provide an even stronger binding defect site than a V$_O$. DFT calculations show that the metal atoms bind between the oxygen adatom and a surface "bridging" oxygen atom in a 2-fold coordination. Note that Ag has a 2-fold coordination in the Ag$_2$O bulk oxide. For Ag, a bond strength of 1.35 eV was obtained with a diffusion barrier of approximately 0.95 eV.

STM investigations of Ag on r-TiO$_2$(110) find no evidence for stable Ag single atoms. Indeed, Ag clusters seem to be formed already at 100 K[44] even if the surface is bombarded by Ar to induce additional surface defects. Cu immediately forms clusters on TiO$_2$ upon deposition at room temperature,[45] but oxygen exposure leads to the formation of 2D islands.

Very little experimental data exists for single Cu atoms on rutile TiO$_2$(110). DFT-based calculations predict that Cu binds in a bridging position on oxygen terminal rows on the stoichiometric surface with a binding energy in the range −230 to −265 kJ mol$^{-1}$.[46,47] This is greater than the adsorption energy at oxygen vacancy sites, suggesting a difference in behavior to Au. In general,[41] it seems that noble metal atoms can bind as cations on unsaturated oxygen atom sites or as anions on unsaturated Ti atom sites. Giordano et al[47] comment that the stronger binding of Cu and Ag over Au stems from the lower ionization potential of these elements, which more readily give away the outermost *s* electron and become cationic. Au, on the other hand, is extremely electronegative, and readily accepts electrons from surface V$_O$ sites leading to a stronger bond in this location. This trend can also be viewed through the absolute Lewis acid hardness.[48] Since Au (Lewis hardness 3.5) is harder than Cu (3.25), it is expected to bind more strongly than Cu on reduced rutile TiO$_2$, where it acts as a Lewis acid in its interaction with a V$_O$. The opposite is then true on stoichiometric TiO$_2$, where Au acts as a Lewis base in its interaction with bridging oxygen atoms. The latter concept is useful in understanding how doping the TiO$_2$(110) surface affects noble metal atom binding. The presence of Cl defects makes the surface more basic, and thus electron transfer more difficult, which reduces the Au bond strength accordingly[48] and explains why the presence of Cl leads to enhanced sintering of Au clusters on TiO$_2$(110).[49] Interestingly, recent DFT calculations have shown that iodine doping might be able to enhance the binding of Ag, Cu and Pd adatoms on TiO$_2$(110). While no additional electron transfer was found that could explain a significant increase in the adsorption energies for the metal at the O$_{2c}$ site in comparison to other halogen doped surfaces, the authors note significant hybridization between the metal, O$_{2c}$, and I states. This suggests that I forms covalent bonds to the metals through the TiO$_2$ surface.[48] It will be fascinating if this stabilization could be verified by experiment, as to date the doping of the metal oxide is a little studied strategy to affect the stabilization of metal adatoms.

## 2.1.2     Ni, Pd, Pt on TiO$_2$(110)

A room temperature STM study of Ni adsorption on r-TiO$_2$(110) reported 3D Ni clusters on the terrace sites even at extremely low coverage.[50] Interestingly, annealing these Ni clusters in O$_2$ resulted in their breakup, and the formation of 2D oxidized Ni islands.[45] An EXAFS study published the same year[51] by the same group suggested that Ni atoms preferentially occupy step edges in a Ti substitutional site. Several DFT studies calculate the favored position for hypothetical Ni adatoms on r-TiO$_2$(110),[52,53] with the most modern calculations favoring adsorption directly above an in-plane oxygen atom. It seems likely that this could only be stabilized at low temperature.

Pd adsorption was studied by STM by Goodman and coworkers, who reported that the smallest stable species was a Pd dimer,[54] and found a distinct preference for Pd clusters to occupy the step edges. There have been several theoretical studies,[55] which find that Pd atoms would, in principle, be most stable in a V$_O$ site. On a stoichiometric surface, the Pd adatom adsorbs in a similar way to Ni,[56] described above, in a hollow site between bridging and in-plane O atoms. The difference to the site directly above the in-plane O atom is negligible, and the diffusion barrier for diffusion along the direction of the bridging oxygen rows is less than 0.05 eV. This well explains why dimers, and larger clusters, rapidly form in the Pd/TiO$_2$(110) system.

Onishi and coworkers were one of the first groups to intentionally image Pt atoms adsorbed on r-TiO$_2$(110).[57] They identified three adsorption sites using noncontact AFM: atop the 5-fold Ti atoms, atop the bridging O rows, and in bridging $V_O$ sites. Only the atoms in the $V_O$ sites were immobile during room temperature imaging. Subsequent studies from Perez et al.[58,59] using ncAFM and DFT calculations suggested that the mobile species observed on the bridging oxygen rows by Onishi were most likely OH groups, which form through reaction of residual water with $V_O$.

Wang and coworkers studied the adsorption of 0.01 ML Pt atoms on r-TiO$_2$(110) using STM, and compared their data to theoretical calculations (see Figure 3).[60] They found Pt atoms to adsorb solely at the $V_O$ sites at 80 K, with no evidence for occupation of the Ti$_{5c}$ sites or the bridging oxygen rows. Accompanying theoretical calculations suggest the Pt atoms trapped by a $V_O$ protrude higher than the neighboring bridging oxygen atoms, and bond to two six-fold coordinated Ti atoms forming a symmetric Ti-Pt-Ti configuration along the [001] direction [Figure 3(d)]. The excess electrons from the vacancy accumulate at the Pt atom leading to a Bader charge of 0.9 e$^-$. Note that this differs from the general idea that the metal adatoms prefer substitutional cation sites on oxide materials and are positively charged. Exposure to CO led to Pt-CO species situated at the $V_O$ sites, but the CO leans toward one of the neighboring Ti rows [Figure 3(e-f)]. While the β-Pt-CO configuration was found to be the more stable by almost half an electron volt, the α-Pt-CO seemingly fits better to what is observed in STM. The authors also measured STM after O$_2$ exposure, and found evidence for molecular adsorption at the Pt sites at 78 K. Unfortunately, no study of the thermal stability of the adatoms was conducted as part of this work, so it was not clear if Pt situated at $V_O$ sites would be stable at non-cryogenic temperatures. Extremely recently, the present authors group performed a study using STM, and confirmed that Pt atoms are indeed stable at $V_O$ sites at room temperature.[61]

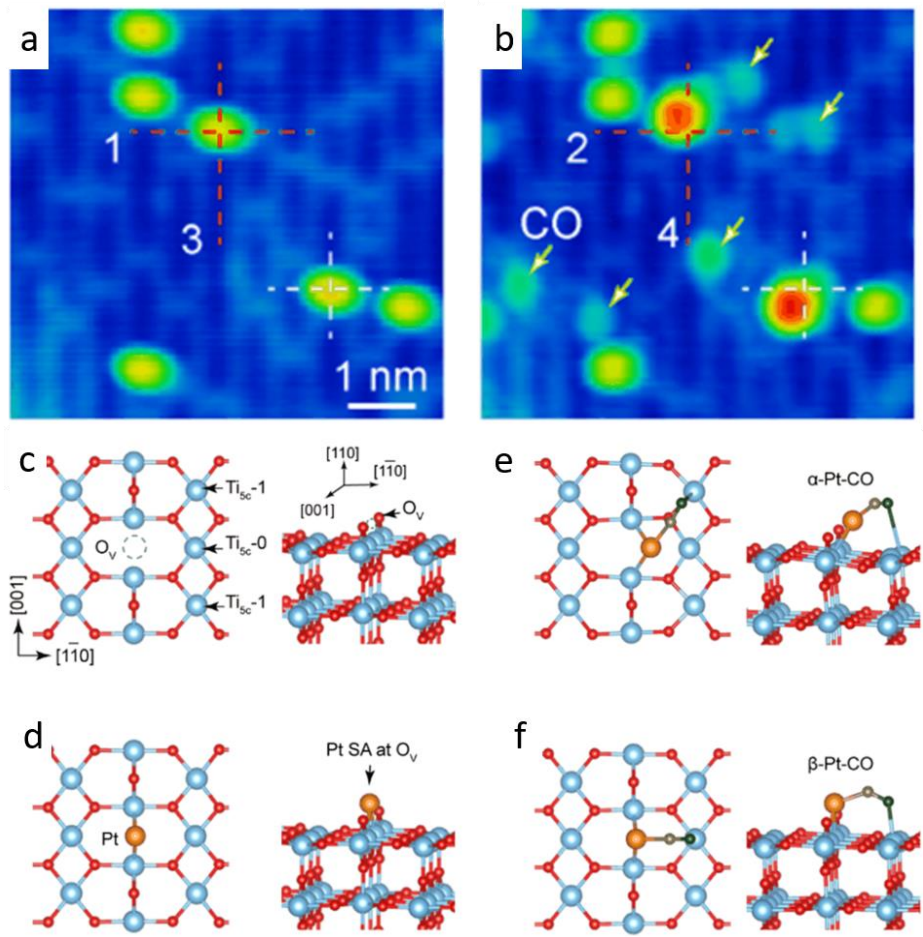

**Figure 3:** STM images at 80 K showing 0.01 ML Pt adsorbed on the r-TiO$_2$(110) surface before (a) and after (b) exposure to CO. The red and white crosses in panel (a) show the position of a V$_O$ on the surface prior to Pt deposition. In (b), the Pt related protrusions become larger (red) due to the adsorption of CO and move away from the vacancy site. This suggests the CO tilts towards the neighboring Ti row. The yellow arrows in (b) highlight CO adsorbed on the Ti rows. Panels (c-f) show DFT models of the clean surface (c), the Pt (orange) adsorbed in the V$_O$ site (d) and two variants of the Pt-CO species that were calculated (e-f). The CO carbon and oxygen atom are drawn in grey and black, respectively. Figure reprinted with permission from ref. 60. Copyright 2017 AIP publishing.

Room temperature ncAFM experiments of Pt adsorption on a h-TiO$_2$(110) surface (i.e. all V$_O$ sites removed by reaction with water) were performed by Pérez and coworkers.[59] Following the deposition of Pt, large but uniform protrusions were observed atop the $Ti_{5c}$ rows, which the authors suggest are due to Pt atoms mobile around a single $Ti_{5c}$ site. Theoretical calculations suggest that the Pt atom can interact with a number of different nearby surface oxygen sites. The features assigned as Pt atoms appear as the brightest protrusions on the surface irrespective of the AFM imaging mode (i.e. the nature of the tip termination), partly because they interact strongly with the AFM tip, a feature which allows to easily distinguish them from the surface OH groups in ncAFM studies. In a follow up paper,[58] the same group also acquired KPFM images of the system in which the species assigned as Pt atoms appear significantly darker than the surrounding $TiO_2$ support. This suggests charge transfer from the Pt atom into the surface, consistent with their simulations. It is nevertheless surprising that Pt adatoms would be stable on the Ti rows, given that Pt atoms are seemingly able to diffuse at 80 K and find the available $V_O$ sites on the r-$TiO_2$(110) surface. It is of course possible that the presence of the hydroxyl species on the surface affects the mobility of the Pt species, and it would be interesting to know the diffusion barrier along the Ti rows in this situation from DFT.

This conclusion of stable Pt on the h-$TiO_2$(110) surface seems at odds with the results of Wendt and coworkers,[62] who compared Pt nanoparticle nucleation on the r-$TiO_2$(110), o-$TiO_2$(110) and h-$TiO_2$(110) surfaces at 90-110 K, 300 K, and after annealing at 800 K. The location of Pt atoms was not the central focus of the study, but nonetheless the observation of the nanoparticles and accompanying theory sheds light on the differing behavior on these three surfaces. At low temperature, Pt nanoparticles were already observed at a coverage of 0.025 ML in all cases, with no apparent preference for step edge or terrace sites. Similar results were found at room temperature for the reduced and oxidized surfaces, but the surface with hydroxyl groups exhibited significantly larger nanoparticles more frequently found at step edges. DFT calculations predicted low diffusion barriers of 0.33 eV and 0.39 eV on the stoichiometric and reduced surface, respectively, suggesting that the as deposited metal would diffuse even at 150 K. $Pt_1$ adatoms were considered to be trapped at a point $V_O$ (−4.28 eV vs. a gas phase Pt atom), while an oxygen atom can also function as a trapping site. Here, the Pt atom is bound between the oxygen adatom and a surface bridging oxygen in a 2-fold configuration. Interestingly, the authors predict that the major difference with the h-$TiO_2$(110) surface is the immobility of the surface defect, as the Pt atom can diffuse as a $Pt_1H$ entity with a barrier less than 0.5 eV. The $V_O$ and O adatom, in contrast, do not diffuse at these temperatures. After annealing at 800 K, no difference is observed because trapping is not effective at this temperature.

One further possibility for a Pt adsorption site at the $TiO_2$(110) surface was proposed by Chang et al in 2014 on the basis of high-angle annular dark-field (HAADF) STEM data.[63] The authors suggest that Pt atoms can occupy 5 different sites at room temperature, with most residing in oxygen vacancy sites located within the Ti-O basal plane (rather than the bridging oxygen vacancy row). The authors rationalized this surprising finding using DFT calculations, which show that while the formation energy of the bridging oxygen vacancy is lower than the in-plane vacancy by 0.15 eV, the total energy gained by placing Pt in an in-plane vacancy is greater. Later, they performed DFT+$U$ calculations that suggested that in-plane oxygen vacancies could coexist with bridging oxygen vacancies in a dilute defect regime. Nevertheless, it is important to note that this defect and/or metal occupation has not been observed in scanning probe studies in UHV, and it is possible that the preparation of the surface by in-air annealing or the electron beam utilized in the experiments might have had an effect. It is known, for example, that annealing a $TiO_2$(110) sample in $O_2$ results in an irregular termination due to the oxidation of Ti interstitials at the surface.[64]

As mentioned above, the Thornton group[39] published a paper in which it was clearly demonstrated using STM that Au atoms occupy bridging oxygen vacancies, not in-plane vacancies. The seemingly incontrovertible proof of the Au position was addressed in a second HAADF-STEM study,[65] which agreed that Au indeed occupies a regular bridging $V_O$ site. The authors finally concluded that strong hybridization between Pt-5$d$6$s$ / $O_{br}$-2$p$ orbitals and Pt-5$d$ / $Ti_{5c}$-3$d$ orbitals are responsible for a 1 eV gain in adsorption energy for occupation of a bridging oxygen vacancy over an in-plane vacancy, despite charge transfer clearly being higher in the latter case. It would certainly be very nice to see a similar study to that performed by Thornton and coworkers for Pt atoms, to ascertain once and for all if in-plane $V_O$ occupation is possible.

## 2.1.3 Co, Rh, Ir on TiO$_2$(110)

No studies of Ir or Rh adsorption on single crystal TiO$_2$(110) in a low coverage regime could be found. However, a recent study of Rh atoms on rutile TiO$_2$ nanoparticles is accompanied by a thorough theoretical analysis, with predictions made for what might be observed.[66] Figure 4 shows that Rh atoms prefer to substitute a Ti cation in a 6-fold surface site under oxidizing conditions (beneath the bridging oxygen row, black line in Figure 4). This is consistent with the experiment in so far as no CO can be adsorbed to conduct an IRAS experiment after the sample was annealed in an oxygen atmosphere. As mentioned above, annealing a TiO$_2$(110) sample in oxygen leads to the oxidation of Ti interstitials at the surface and the growth of new material, *i.e.* growth of new TiO$_2$ at step edges or as new terraces.[64] At O$_2$ chemical potentials between −1.7 eV and −2.5 eV, a V$_O$ is predicted to form above the Rh cation, which reduces its coordination to 5-fold (green line in Figure 4). Interestingly, in extremely reducing conditions, the Rh atom is found to prefer a "supported geometry", which is a site above an in-plane oxygen atom in which the Rh is coordinated to both Ti and O atoms from the surrounding surface (blue line in Figure 4). Note that viewed from above (as in a STEM experiment), this site could look very much like adsorption in an in-plane V$_O$. The authors go on to discuss the influence of H$_2$ and CO atmospheres, and show that adatom geometries can compete with the stable substitutional ones in realistic catalytic conditions. Finally, the authors show that the undercoordinated Rh configuration can coordinate two CO molecules, while the substitutional geometry cannot. IRAS experiments conducted in a CO rich atmosphere clearly show the signature of the Rh(CO)$_2$ dicarbonyl, consistent with the theoretical predictions. It is important to recognize, however, that the experiments were performed on nanoparticles with a 30 nm diameter, while the calculations utilized terrace sites on a TiO$_2$(110) slab using periodic boundary conditions. The direct comparison is thus not ideal, but does nevertheless serve to validate the theory that the Rh adatoms change their adsorption site depending on the environment.

Very recently, the present authors group studied Rh adsorption on r-TiO$_2$(110) using room temperature STM in UHV conditions. Isolated Rh atoms were observed after deposition at 100 K, with no preference for the V$_O$ sites. Annealing to 150 K was already sufficient to sinter the adatoms into small clusters, which grew larger upon annealing to 300 K. Consequently, it seems that Rh diffusion is facile on TiO$_2$(110), and this will prevent the stabilization of Rh$_1$ species at temperatures relevant for SAC.[61]

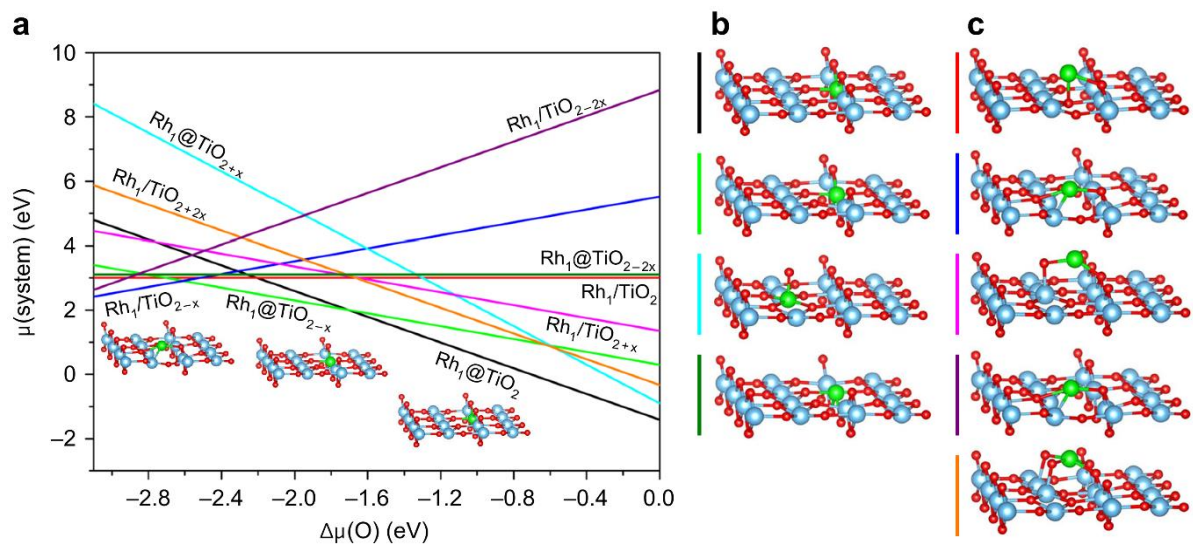

**Figure 4:** Atomistic thermodynamics for substitutional (@) and supported (/) Rh geometries at the TiO$_2$(110) surface. **a** Relative stability as a function of oxygen chemical potential Δ$\mu$(O). **b, c** The optimal structures for substitutional (**b**) and supported (**c**) Rh SAs on the considered TiO$_2$ surfaces. The lowest-energy structures are also shown as insets in **a**. Color code: O—red; Ti—blue; Rh—green. Figure adapted with permission from Ref. 66. Copyright 2019, Springer Nature under CC-BY license (https://creativecommons.org/licenses/by/4.0/).

Chen and coworkers[67] compared the nucleation and growth of Co particles to other metals (Au, Ni and Pt). In similar conditions, cluster sizes increase in the order of Co < Pt < Ni < Au, suggesting Co has a stronger interaction with the TiO$_2$(110) surface.

### 2.1.4   Fe on TiO$_2$(110)

STM images of Fe at the TiO$_2$(110) surface suggest that Fe can be stabilized at the V$_O$ sites.[68] Small clusters are also imaged after room temperature Fe deposition, and, while significant sintering occurred after heating to 473 K, the features assigned to single Fe atoms survived

## 2.2 Anatase TiO$_2$

Most, but not all, surface science studies of anatase TiO$_2$ focus on the most stable (101) surface. After preparation in UHV the surface exhibits a sawtooth-like surface termination that exposes a mixture of fully coordinated and undercoordinated Ti and O atoms[69] (see Figure 5). The surface is well characterized in terms of its structure[70,71] and molecular adsorption,[72-74] and is thus in principle suitable as a model system for SAC purposes. One of the key differences between anatase TiO$_2$(101) and rutile TiO$_2$(110) in that surface oxygen vacancies are not stable on the former surface. Even if such species are created artificially (by low energy electron bombardment), they diffuse into the bulk already at low temperature and a stoichiometric surface is recovered.[74,75] Thus, such sites would not be expected to be available to stabilize metal atoms under ambient conditions.

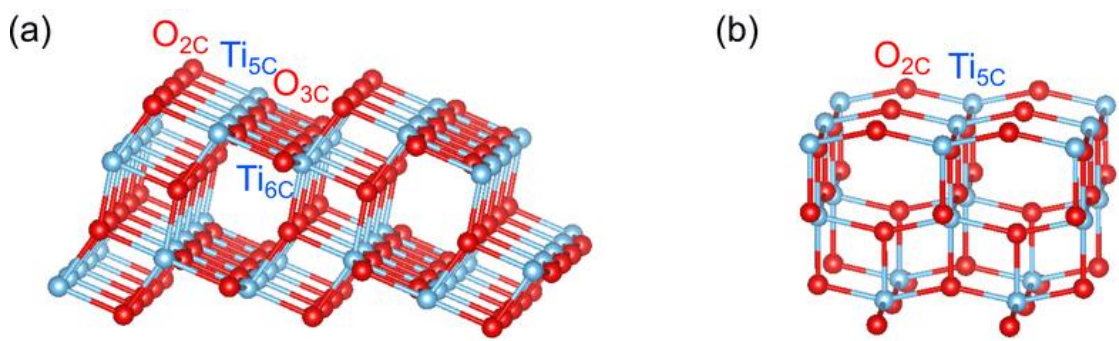

**Figure 5**: Bulk-terminated surfaces of (a) anatase (101) and (b) anatase (001). Figure adapted from ref. [76]. Copyright 2018 MPDI under CC-BY license (https://creativecommons.org/licenses/by/4.0/).

### 2.2.1    Au, Pt on anatase TiO$_2$

Experimental surface science studies of metal adsorption on anatase $TiO_2$(101) are limited, and mostly predate a particular interest in obtaining isolated single atoms. Diebold, Selloni and coworkers studied Pt and Au evaporated onto anatase $TiO_2$(101) using STM and found that both systems sinter already at low coverage resulting in nanoparticles.[77] Au was found to interact weakly with the surface (adsorption energy just 0.25 eV), with adatoms computed to be most stable directly above a $Ti_{5c}$ atom. The weak interaction with Au leads to large nanoparticles in experiment, and a strong preference for the step edge was observed in room temperature STM images. When the support was irradiated with electrons to create $V_O$ sites prior to Au deposition, smaller clusters were observed on the terraces, suggesting nanoparticle nucleation occurred at the $V_O$ sites. This is consistent with the much larger adsorption energy 3.16 eV computed for $Au_1$ at a $V_O$ site.

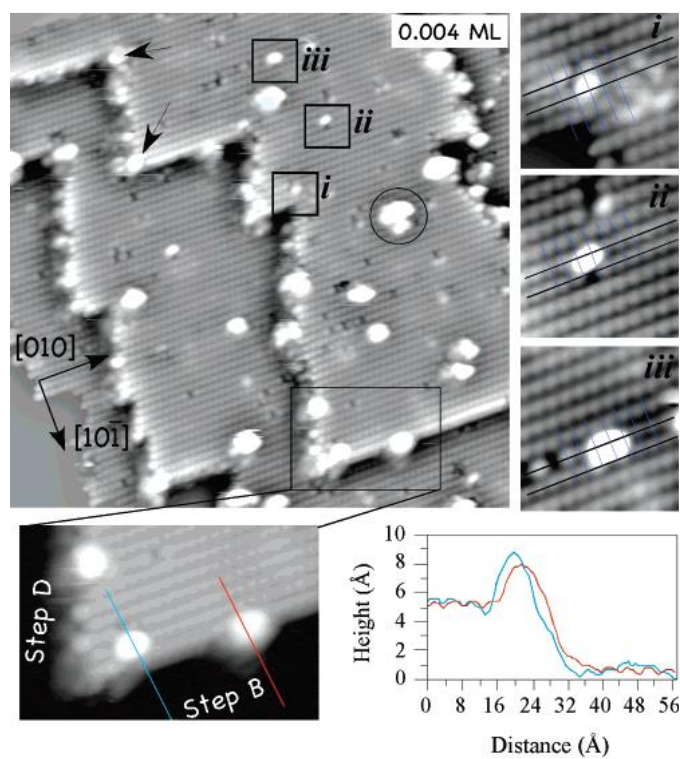

**Figure 6**: STM images of the anatase(101) surface after deposition of 0.004 ML Pt. The inset and line profiles at the bottom show that Pt clusters at step edges are mostly located at the upper terraces, with kink sites are preferred nucleation sites (see black arrows). The clusters in the insets (i-iii) were tentatively assigned as a (i) Pt monomer, (ii) Pt dimer, and (iii) Pt trimer. Reprinted with permission from ref. [77]. Copyright 2008 American Chemical Society.

In the case of Pt, clusters nucleated both on the terrace and at the step edges (see Figure 6), and the observation of atom-sized protrusions was interpreted as Pt single atoms (Figure 6, inset i). However, the site of these species atop surface $O_{2c}$ atoms disagreed with the DFT+$U$ predicted site, and the protrusions appear similar to those observed subsequently for adsorbed water. Slightly larger protrusions were tentatively assigned to dimers and trimers. Theoretical calculations suggest that the optimal location for a Pt atom is in between two $O_{2c}$ sites, close to two $Ti_{5c}$ and two $Ti_{6c}$ surface atoms. The average Pt−$O_{2c}$ and Pt−Ti distances were ∼2.04 and 2.78 Å, respectively. The $Pt_1$ adsorption energy was found to be fairly strong at 2.2 eV, but no diffusion barrier was computed. Pt deposition on a reduced anatase $TiO_2$(101) surface with $O_{2c}$ vacancies results in smaller Pt clusters, and the accompanying DFT calculations suggest Pt would have a much stronger adsorption energy (4.7 eV) at the $V_O$ site.

More recently, an extremely thorough computational analysis of the $Pt_1$/anatase $TiO_2$(101) system was performed by Pacchioni's group to accompany experiments performed on a powder catalyst. Interestingly, the $Pt_1$/$TiO_2$ catalyst was prepared with an extremely low loading such that most particles of the support would contain just one atom, thus avoiding the possibility for agglomeration into clusters.[78-81] The system was characterized by IRAS as a function of temperature, allowing both the CO-stretch signature and CO binding energy of the Pt species to be simultaneously determined. These parameters can be extracted from DFT-based models of the system, and were used to test the validity of the results. Occupation of different sites clearly makes a large difference to the properties of the metal atom, and through this, the CO binding properties. However, none of the simple candidate positions for Pt on an anatase $TiO_2$(101) surface (adatom bound to surface oxygen, substitutional cation site, occupying an oxygen vacancy, etc.) reproduced the CO-stretch frequencies and binding energies observed experimentally. Ultimately, the authors proposed that additional coordination to surface OH groups could produce good agreement. Such species are omnipresent on metal oxide surfaces in realistic conditions, but are almost never considered in SAC calculations. It is important to note however that this does not constitute proof that this configuration is actually present on the samples. Nevertheless, this work clearly shows that the use of simplistic models is not sufficient to capture the complexity of real SAC systems.

## 2.2.2 Rh on anatase $TiO_2(101)$

There is a dearth of experimental surface science studies of Rh, Ir or Ni atoms on anatase $TiO_2(101)$. However, Christopher and Pacchioni performed an important study that shows how the typical pretreatments performed in SAC affect the properties of a $Rh/TiO_2$ powder system.[80] Following a standard solution-based synthesis procedure, the catalyst was heated in $O_2$ at 350 °C, ostensibly to remove the ligands. The resulting material did not adsorb CO at room temperature at all, suggesting that the Rh does not reside at the surface of the $TiO_2$ nanoparticles. The authors assume that Rh atoms substitute cations in the bulk $TiO_2$ lattice, which seems reasonable (note, the present authors group has shown this phenomenon directly for $Fe_3O_4$ and $\alpha\text{-}Fe_2O_3$).[82,83] In any case, reducing the sample by heating in either $H_2$ or CO atmosphere modifies the catalyst such that CO adsorption becomes possible at Rh sites, and the authors clearly see the IRAS signature of the Rh dicarbonyl. This suggests the presence of isolated Rh atoms. While the room temperature IRAS signature was the same whether CO or $H_2$ was used as the reductant, the thermal evolution of the systems was clearly different. Following CO pretreatment, both CO molecules desorb simultaneously from the dicarbonyl at 240 °C, whereas following $H_2$ reduction, some fraction of the species form a distinct intermediate. It makes sense that the surface reduced in $H_2$ might exhibit hydroxyl groups, and DFT calculations were indeed able to show that a $Rh(OH)(CO)_2$ could produce the properties observed in experiment. The key takeaways from this study, however, are that the pretreatment makes a significant difference to the resulting properties of the catalyst, and that room temperature IRAS alone is not sufficient to distinguish the different species on the surface.

## 2.3 Conclusions

Overall, TiO$_2$ remains an interesting model system for SAC because of the rich interaction of adatoms with its various defects. The surface science experiments performed on rutile TiO$_2$(110) to date reveal that V$_O$'s are the most stable sites to stabilize electronegative metal atoms. V$_O$ sites react strongly with water, however, and thus will not be present at the surface during wet synthesis of real systems. Moreover, the calcination treatment typically employed in the synthesis of a real SAC system is oxidizing, and thus will not lead to the generation of V$_O$ sites. It is likely, however, that V$_O$ sites will be created during activation of the catalyst (which generally involves heating in a reducing atmosphere), and that metal atoms previously stabilized in other sites, or in clusters, could migrate to V$_O$ sites, and remain stable there in a reactive environment. As such, it is interesting to consider what catalytic properties the resulting negatively charged metal adatoms might have, and there are several computational investigations where Au$_1$/r-TiO$_2$(110), for example, is predicted to be a good catalyst system.[84-87] As yet, there have not been experimental investigations to confirm these predictions. More fundamentally, it would be interesting to study how a UHV prepared TiO$_2$(110) surface is modified by realistic calcination and reduction treatments, and if isolated Au or Pt adatoms reside in V$_O$ sites afterwards. Other metals of interest, for example Rh, which do not preferentially occupy V$_O$ sites, seem to diffuse and sinter too readily to be promising SAC systems, unless they could be stabilized by coordination to additional ligands.

The studies to date on anatase (101) provide little evidence for stable metal adatoms. In contrast to rutile (110), V$_O$'s are not stable on the surface, but it is possible that they could be formed during reduction and rapidly occupied by diffusing metal adatoms. It is important to note that the experiments performed by the Christopher group that suggest the Pt/anatase system to be active for CO oxidation utilized a very low loading that ensured each anatase particle supported only one metal adatom. As such the sintering observed at room temperature in surface science experiments was avoided. It would of course be possible to investigate the most stable site at cryogenic temperatures, but combining structural studies of reactivity will inevitably be difficult. In the present authors opinion, it seems that the TiO$_2$ surfaces primarily studied so far are not ideal model systems for SAC research. It is the stable nature of these surfaces that makes them somewhat unreactive, however, and it would be interesting to learn if the different structures and potential coordinations presented by other less stable facets could enhance adatom stabilization.

## 3. Iron Oxides (FeO$_x$)

Much of the pioneering work on single atom catalysis from the group of Zhang and coworkers utilizes iron oxide as the support.[29,88,89] In their 2011 *Nature Chemistry* paper,[90] for example, it was shown that Pt atoms bound to iron oxide particles are cationic, and that these species catalyze CO oxidation and preferential oxidation of CO (PROX) as efficiently as Pt nanoparticles. The iron-oxide support employed by Zhang and coworkers is nominally α-Fe$_2$O$_3$, but the FeO$_x$ notation acknowledges that the surface is likely reduced following activation (i.e. heating in a reducing atmosphere). Iron has three stable oxides (FeO, Fe$_3$O$_4$, Fe$_2$O$_3$), but intermediate stoichiometries are possible and it is relatively easy to change between them, particularly at the surface.[91] Depending on the conditions, it is possible that a hematite (α-Fe$_2$O$_3$) surface can even change phase locally to form Fe$_3$O$_4$, even if the bulk remains Fe$_2$O$_3$. Nevertheless, for simplicity, the theoretical calculations accompanying studies of FeO$_x$-based catalysts utilize an idealized α-Fe$_2$O$_3$(0001) structure. However, surface science studies indicate a highly complex surface phase diagram for the α-Fe$_2$O$_3$(0001) facet, and the idealized (1×1) iron- or oxygen-terminations assumed in most DFT studies may be too simplistic. In what follows we briefly discuss what is known about metal adsorption on α-Fe$_2$O$_3$(0001), before discussing recent work on a different low energy facet: α-Fe$_2$O$_3$(1$\bar{1}$02). Following this, we describe what has been learned from what we consider to be the best characterized model system for SAC: Fe$_3$O$_4$(001).

### 3.1 α-Fe$_2$O$_3$(0001)

The (0001) facet of hematite has been studied extensively using the surface science approach, but significant disagreement remains about its possible terminations and especially their respective stability regions.[91] Most studies report iron- or oxygen-terminated bulk truncation models,[92,93] but experimental evidence for a ferryl termination has also been reported.[94,95] More critically, UHV preparation also often results in a complex superstructure with ≈ 4 nm periodicity, generally referred to as the "biphase" termination.[96-99] The nature of this biphase structure remains poorly understood at an atomic scale, and is effectively inaccessible to theory due to the large number of atoms per unit cell. DFT modelling of $FeO_x$-based SAC systems generally assumes one of the bulk-truncated (1×1) terminations. Adatoms are generally placed in three-fold hollow sites of the oxygen-terminated surface,[90] which is equivalent to adatoms substituting iron in the iron-terminated surface. Catalytic activity has been theoretically screened for a range of different metals in this configuration.[28,100] However, the experimental evidence for the adatom site comes mainly from TEM, which shows the single atoms in cation-like positions with respect to the hematite bulk.[90] This makes the assignment of an Fe substitution site plausible, but other explanations are also still possible.

In the surface science literature, there is no experimental evidence for single atoms being stabilized on the α-$Fe_2O_3$(0001) surface. Few deposition experiments have been attempted, likely due to the overall complexity of the surface. In a recent study, Lewandowski et al. deposited Fe and Au on the biphase termination, and showed that both metals initially accumulate in only one of the three distinct regions found in the superstructure.[98] Small gold clusters prepared in this manner have subsequently been shown to be active for CO oxidation.[101] However, even if some single atoms were present in these experiments, they are clearly not stabilized at high loadings, and the biphase likely is not a good representation of nanoparticle catalyst surfaces.

Overall, there remains a significant experimental gap between theoretical models assuming a simple (1×1) bulk truncation and actual catalysts, where the surface termination is unknown. However, both the variety of reported terminations and the difficulty in theoretical treatment of the "biphase" make the α-$Fe_2O_3$(0001) an extremely challenging and ultimately unattractive model system for surface science.

## 3.2    α-$Fe_2O_3$(1$\bar{1}$02)

The α-Fe$_2$O$_3$(1$\bar{1}$02) surface is non-polar, and exists in a (1×1) bulk truncation after UHV preparation, although a reduced (2×1) termination is formed in reducing conditions.[102] This well-defined structure is in principle ideal for SAC studies, and the first such investigation were published recently. Upon deposition at room temperature, Rh was shown to form clusters (Figure 7).[83] However, when the sample is heated, the clusters disappear, and the Rh atoms become incorporated in the lattice of the support. Based on a combination of STM, low energy ion scattering (LEIS) and DFT, it was concluded that the redispersed Rh atoms are located in the immediate subsurface layer.[83] Of course, this means that the Rh atoms will be inaccessible for reactants, and thus this system will likely not be active as a single atom catalyst. However, it illustrates that great care must be taken in identifying single atoms as active sites, as the subsurface Rh atoms may easily be identified as single atoms in cation-like sites at the surface by TEM.

While Rh on α-Fe$_2$O$_3$(1$\bar{1}$02) forms clusters at room temperature in UHV, it can be stabilized by co-adsorption with water, which is stable on this surface up to 345 K.[103] This results in Rh(OH)$_2$ species (Figure 8), which are mobile at room temperature, but do not agglomerate to clusters.[104] This result is interesting because water and hydroxyl groups are omnipresent on metal oxides in atmospheric conditions, so their presence should be generally taken into account in the modeling of oxide-supported SAC systems.

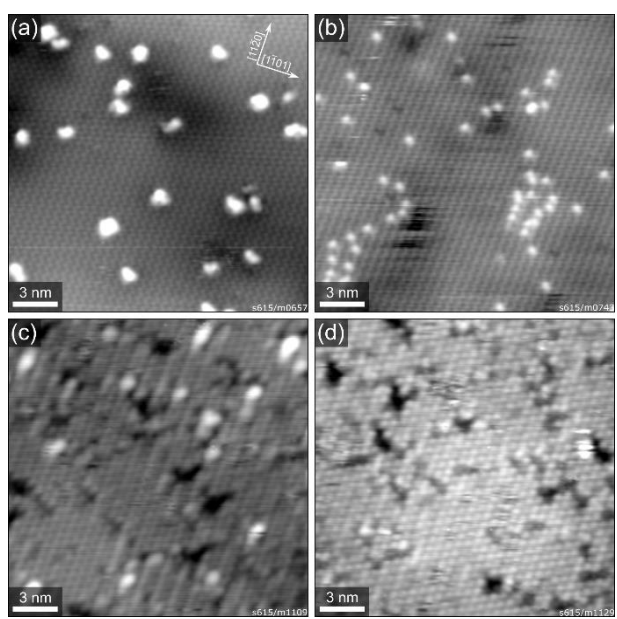

**Figure 7:** STM images of 0.025 ML Rh on α-Fe$_2$O$_3$(1$\bar{1}$02). (a) 0.025 ML Rh as deposited on the clean α-Fe$_2$O$_3$(1$\bar{1}$02)-(1 × 1) surface at room temperature ($U_{sample}$ = +3 V, $I_{tunnel}$ = 0.3 nA) and (b) after annealing at 500 °C for 15 minutes in UHV ($U_{sample}$ = −2.8 V, $I_{tunnel}$ = 0.1 nA). (c) 0.025 ML Rh as deposited on the clean α-Fe$_2$O$_3$(1$\bar{1}$02)-(2 × 1) surface ($U_{sample}$ = −3 V, $I_{tunnel}$ = 0.1 nA) and (d) after annealing at 300 °C for 10 minutes in UHV ($U_{sample}$ = −2.8 V, $I_{tunnel}$ = 0.1 nA). Reproduced with permission from ref. [83]. Copyright 2021 Wiley VCH under CC-BY license (https://creativecommons.org/licenses/by/4.0/).

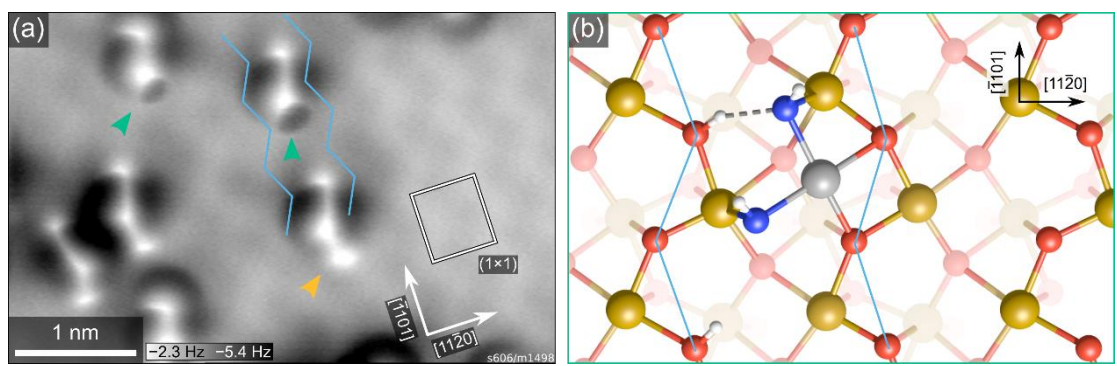

**Figure 8:** Rh on α-Fe$_2$O$_3$(1$\bar{1}$02) stabilized by co-adsorbed water. (a) ncAFM image acquired at liquid He temperature of 0.05 ML Rh on α-Fe$_2$O$_3$(1$\bar{1}$02), deposited at room temperature in a partial pressure of 2 × 10$^{-8}$ mbar H$_2$O, then heated to 80 °C to desorb all water not coordinated to Rh. (b) Schematic model (top view) for the features indicated by green arrows in panel (a). A Rh adatom (grey) is stabilized by two OH groups (O$_{water}$ in blue, hydrogen in white). The zigzag rows of surface oxygen are marked in blue in both panels. The orange arrow highlights a third protrusion that is sometimes present, which was attributed to an additional water molecule atop a surface Fe and hydrogen bonded to one of the OH groups. Reproduced with permission from ref. [104]. Copyright 2022 American Chemical Society.

## 3.3   Fe$_3$O$_4$(001)

Fe$_3$O$_4$(001) is an ideal model system to study isolated adatoms under UHV conditions. Following preparation by Ar$^+$ sputter/anneal cycles in UHV, the surface exhibits a (√2×√2)R45° [also known as

c(2×2)] periodicity. STM images reveal undulating rows of Fe atoms running in the [110] directions, consistent with a termination at the plane containing both octahedrally coordinated Fe and oxygen (see Figure 9). The surface layer is bulk-like in terms of stoichiometry, but is distorted due to an ordered array of cation vacancies and interstitials in the immediate subsurface. The proposed structure was confirmed by a combination of quantitative LEED[105] and DFT+$U$ calculations, as well as SXRD[106] measurements. Crucially for our purposes here, the reconstruction has been shown to stabilize ordered arrays of metal adatoms of almost any variety,[91] and is thus an ideal model system to study fundamental processes in SAC. In what follows, we summarize the main results from almost 10 years of work on this surface. However, the focus will be on publications post 2015, as an extensive summary of prior work already exists as part of a review of iron oxide surfaces.[91]

### 3.3.1    Cu, Ag, Au on $Fe_3O_4$(001)

The stabilization of metal adatoms on $Fe_3O_4$(001) was first reported for Au in 2012.[107] It was shown that Au adatoms adsorb midway between the rows of Fe atoms imaged in STM. This location is consistent with 2-fold coordination to surface oxygen atoms (see Figures 9 and 10), in a site which is essentially where the next tetrahedrally coordinated Fe atom would be if the spinel structure were continued outward. The Au atoms were stationary in STM movies at room temperature, and remained stable against thermal sintering to temperatures as high as 700 K. Based on a Monte Carlo simulation, it was suggested that the coverage threshold coincided with the probability for two adatoms to be deposited directly into the same unit cell. The thermal sintering at 700 K (also observed for other metals) seems to be linked to an order-disorder transition that occurs in the surface reconstruction at this temperature. For coverages in excess of $2.1 \times 10^{13}$ cm$^{-2}$, Au clusters were observed to coexist with adatoms after deposition at room temperature.[108,109] STM images of the surface after heating the mixed system were suggestive of a "rolling snowball mechanism" of cluster growth, whereby the clusters diffuse at elevated temperature and pick up otherwise stable adatoms they encounter.

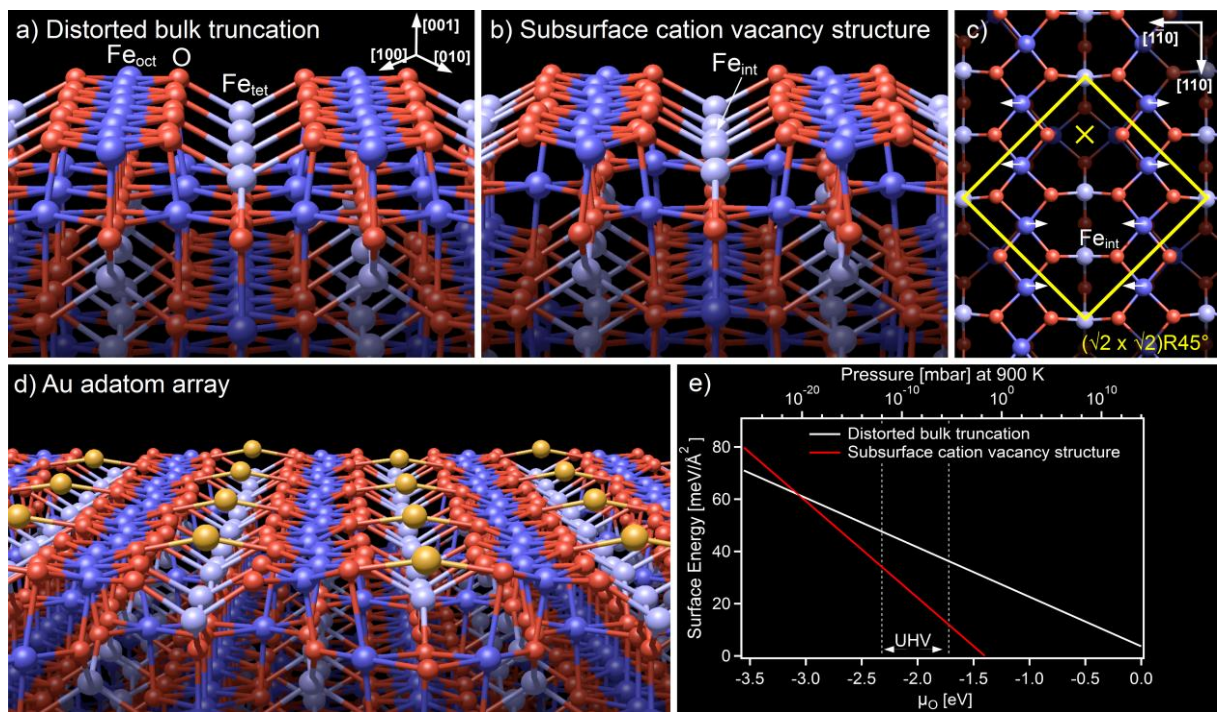

**Figure 9:** a) Schematic model of a bulk-terminated $Fe_3O_4(001)$ surface. b) Schematic model of the surface structure obtained following preparation in ultrahigh vacuum. An interstitial Fe atom with tetrahedral coordination ($Fe_{int}$) replaces two Fe with octahedral coordination ($Fe_{oct}$) from the third layer. c) Top view of the subsurface cation vacancy (SCV) model showing with a yellow × the site in which metal adatoms adsorb on the $Fe_3O_4(001)$ surface. d) DFT+$U$ calculation showing Au adatoms adsorbed on $Fe_3O_4(001)$ at the maximum coverage (defined as 1 monolayer). e) Surface energy versus $O_2$ chemical potential for the two terminations shown in panels a and b, as calculated by DFT. The subsurface cation vacancy structure is more stable in all conditions reachable experimentally. Figure reproduced with permission from ref. [105]. Copyright 2014 American Association for the Advancement of Science.

Experimental observations for Ag were somewhat similar to Au, although the coverage threshold for cluster formation was significantly higher (circa. $7 \times 10^{13}$ cm$^{-2}$, or 50% of the available sites occupied).[110] Using DFT+$U$ calculations, it was shown that the Ag-dimer is unstable compared to two Ag adatoms on $Fe_3O_4(001)$. This was consistent with the observation of Ag adatom mobility at room temperature in STM movies, suggesting that cluster nucleation required a particular minimum size.

Cu adatoms can be stabilized to even higher coverages than Ag and Au,[111] and there is evidence that a second adsorption site can be occupied once the standard configuration (2-fold coordinated to lattice oxygen) becomes saturated. The protrusion appears to be located in the "wide" phase of the surface reconstruction, which puts the adatom directly above the $Fe_{int}$ atom in Figure 9c. It is not clear if the $Fe_{int}$ remains in place, and it seems more likely that it moves to occupy one of the neighboring $Fe_{oct}$ vacancies in the layer below.

DFT+$U$ based calculations suggest that all three noble metal atoms take a 1+ oxidation state when adsorbed on $Fe_3O_4$(001). This makes sense, because all three metals are twofold coordinated to oxygen in their native oxides, where they take a 1+ oxidation state. The assignment is supported by XPS binding energies, which compare well between Cu and Ag adatoms and literature values for the metal oxides.[111] Quantitative normal incidence x-ray standing wave experiments further support the theoretical model of the adsorption geometry, with excellent agreement for the position of the Ag and Cu adatoms with respect to the surface. However, this agreement is contingent in the theoretical lattice parameter being constrained to the experimental value, as expanding the lattice widens the separation of the surface oxygen atoms to which the adatoms bind, causing them to sink lower into the surface to obtain the same binding length. This issue was not encountered with the hybrid functional calculations, primarily because the theoretical lattice parameter comes out very close to the experimental value.[111]

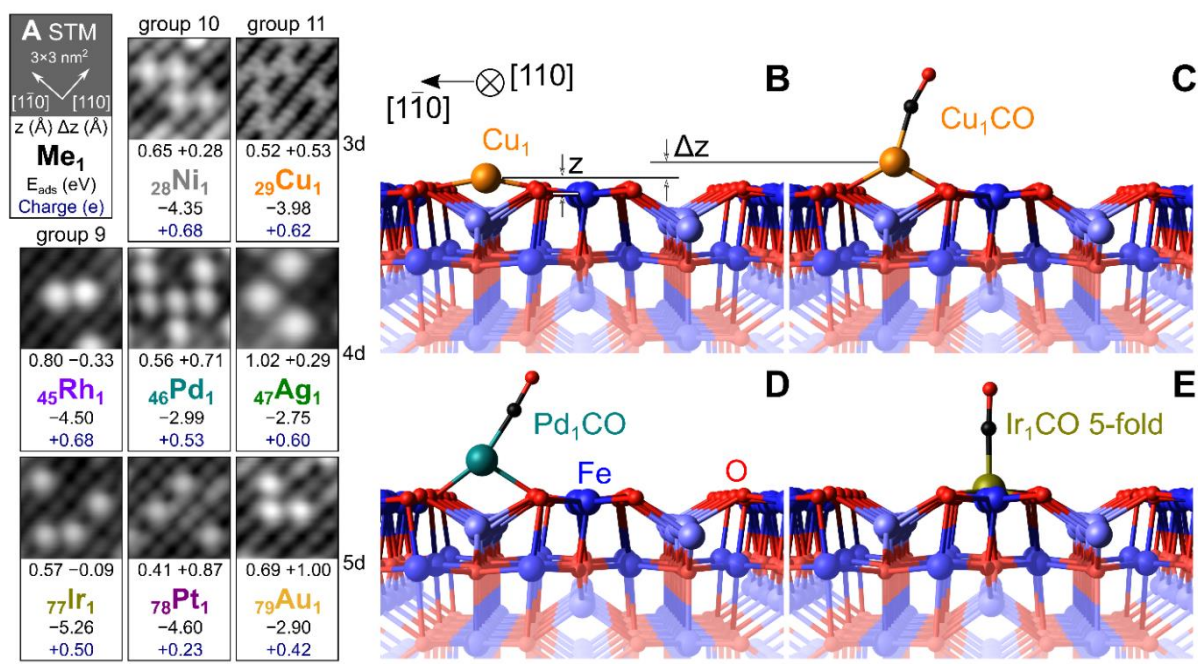

**Figure 10**: (A) Representative STM images ($U_{sample}$ = +1-1.5 V, $I_{tunnel}$ = 0.1-0.3 nA) showing metal adatoms adsorbed between the Fe rows of the $Fe_3O_4$(001) support. Text in each panel indicates the DFT+$U$ derived adsorption energies, Bader charges, and heights of the $Me_1$ adatom ($z$) above the surface Fe atoms in the 2-fold adsorption geometry, as well as the CO-induced vertical displacement ($\Delta z$). (B-D) DFT+$U$ derived minimum energy structure for the 2-fold coordinated $Cu_1$/$Fe_3O_4$(001) adatom before (B) and after (C) adsorption of CO, as well as the $Pd_1CO$ carbonyl (D), which is lifted from the surface. (E) IrCO replaces a 5-fold coordinated surface Fe atom during the TPD ramp, meaning that CO desorption ultimately occurs from the depicted 5-fold $Ir_1$ geometry. Figure adapted from ref. [112]. Copyright 2021 American Association for the Advancement of Science.

### 3.3.2 3d transition metals (Ti, Mn, Co, Ni) on $Fe_3O_4$(001)

All of the 3d transition metals excluding Cu exhibit a similar behavior.[113,114] Upon deposition the adatoms occupy the same 2-fold coordination as shown for Au in Fig. 9d, but this is unstable against incorporation in the $Fe_3O_4$ lattice. This is straightforward to understand because all of these metals form stable solid solution ferrite compounds with the spinel structure ($MeFe_2O_4$, where Me = metal).

The temperature at which incorporation occurs increases from left to right in the periodic table, and a significant proportion of Ti is already incorporated upon deposition at room temperature. This is possible because Fe vacancies exist in the subsurface, and these can be occupied by either the adatom itself, or by an Fe atom displaced from the first layer. If the sample is heated above 700 K then all of the foreign metal diffuses to the bulk of the sample and is undetectable by XPS. Thus, in contrast to the noble metals discussed above, which sinter to clusters, the 3d transition metals are thermodynamically driven to disperse within the oxide.

### 3.3.3     Rh, Ir on $Fe_3O_4(001)$

Rhodium and iridium adsorb in the standard 2-fold coordination on $Fe_3O_4(001)$. These metals are not 2-fold coordinated in their stable bulk oxides however, and thus prefer to incorporate within the $Fe_3O_4$ lattice where octahedral coordination to oxygen can be achieved (see Figure 10E). In both cases, a temperature of at least 500 K is required to initiate the transition. Despite this similarity, Rh ultimately disperses within the support bulk after prolonged heating, whereas Ir leaves the lattice and forms large metallic clusters (see Figure 11). This reflects the higher cohesive energy of Ir metal versus Rh, and the greater oxophilicity of Rh (as judged by the heat of formation of the most stable oxide). For both metals, 6-fold coordination in the subsurface layer is energetically favored over 5-fold coordination in the surface layer. This has implications for SAC, because an atom with 6-fold coordination is inaccessible, and will not interact strongly with reactants.

CO adsorption has been studied on both Rh and Ir adatoms on $Fe_3O_4(001)$ using surface science techniques.[82,112,115] It was found that CO binds strongly, but that adsorption does not lead to destabilization and sintering (as is the case for Pd and Pt). Indeed, the adsorption of a single CO molecule causes the adatom to relax towards the surface, which is linked to the formation of a weak bond to a subsurface oxygen atom. With this, the metal atom takes a pseudo-square-planar environment, akin to the structure found in Ir(I) and Rh(I) complexes. The adsorption of a second CO molecule leads to the formation of a dicarbonyl, and together with the bonds to the support, the metal adatom achieves a square-planar configuration. It is interesting to note that the CO desorption observed in TPD experiments ultimately occurs from a metal atom with 5-fold coordination, because the switch to octahedral coordination in the surface layer occurs with the CO still attached.

For Rh, it was observed that exposure to $O_2$ even at very low pressures leads to destabilization and sintering. The resulting $RhO_x$ clusters are active for CO oxidation, and are difficult to distinguish from Rh adatoms in XPS. This suggests one must be careful assigning $Rh_1$ species on the basis of cationic signature in spectroscopy.

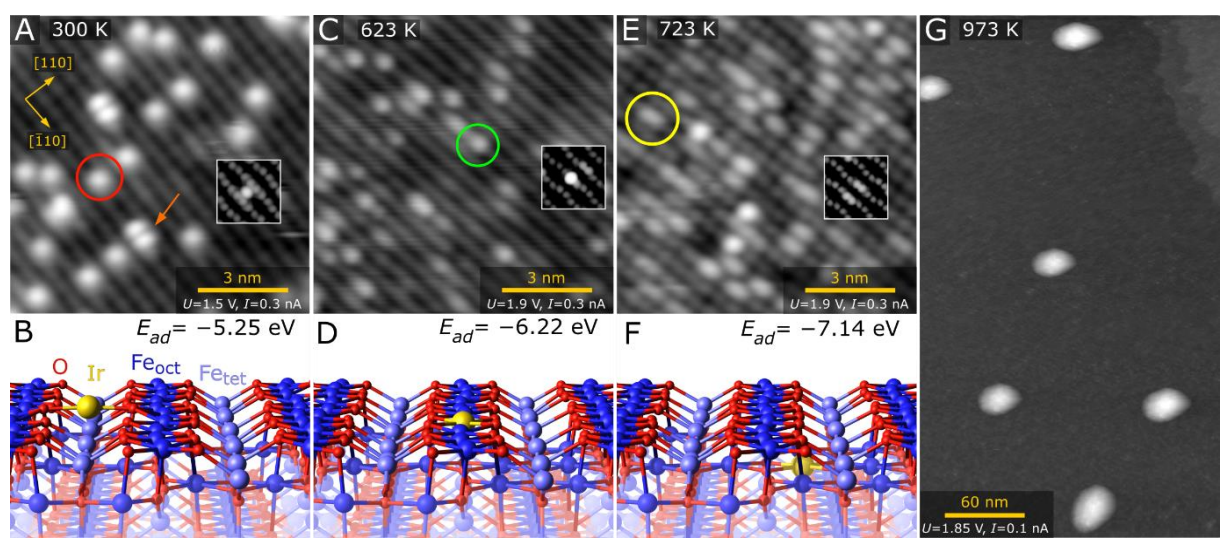

**Figure 11:** (A) $Ir_1$ atoms evaporated directly onto the $Fe_3O_4(001)$ surface at 300 K are imaged as bright protrusions between the Fe rows of the support (red circle in STM image). Double protrusions are metastable $Ir_2$ dimers (orange arrow). (B) DFT-derived minimum-energy structure of the 2-fold coordinated Ir adatom on $Fe_3O_4(001)$. An STM simulation based on this structure is shown as an inset in panel A. (C) After annealing at 623 K, Ir atoms appear as bright protrusions within the Fe row in STM images (green circle). (D) DFT-derived minimum energy structure of the 5-fold coordinated Ir atom incorporated within the $Fe_3O_4(001)$ surface, with the corresponding STM simulation shown as an inset in panel C. (E) At 723 K, some of the bright protrusions within the row are replaced by extended bright protrusions in STM (yellow circle). Some small irregular clusters are also observed. (F) DFT-derived minimum energy structure of the 6-fold coordinated Ir adatom incorporated in the subsurface layer of $Fe_3O_4(001)$. An STM simulation based on this structure is shown as an inset in panel E. (G) Annealing at 973 K leads to formation of metallic Ir clusters with an apparent height of

≈ 3 nm. Figure reproduced with permission from Ref. [115]. Copyright 2019 Wiley-VCH Verlag GmbH & Co. KGaA under CC-BY license (https://creativecommons.org/licenses/by/4.0/).

### 3.3.4 Pt, Pd on $Fe_3O_4(001)$

Pt[116] and Pd[117] both form 2-fold coordinated adatoms on $Fe_3O_4(001)$ upon room temperature deposition. The adatoms are stable to high coverage and temperature (700 K), but both are destabilized by CO, which is omnipresent even in the UHV environment. PdCO and PtCO species are mobile at room temperature, which leads to agglomeration. In the case of Pd,[117] STM movies suggest that PdCO can be immobilized if they encounter a surface hydroxyl group before a second PdCO species, which suggests that $Pd_1$ might be more stable in a realistic environment where the support is hydroxylated. The alternative possibility, which happens often in the movies, is the nucleation of a Pd cluster and further sintering into Pd nanoparticles. In the case of Pt,[116] the most likely event is the formation of a $Pt_2(CO)_2$ species. These are stable and immobile in UHV at room temperature, but break apart if the sample is heated to approximately 500 K. The final state is a mixture of clusters and adatoms, suggesting that further diffusion occurs at this temperature. In addition to destabilization by CO, the same effect has also been reported for Pd adatoms in the presence of methanol.[118]

Very recently,[119] it was shown that the decomposition of the $(PtCO)_2$ dimer is linked to the production of $CO_2$. Based on quantitative, isotopically-labelled TPD measurements and DFT, it was concluded that Pt dimers react with the surface through extraction of oxygen from the lattice. Interestingly, the DFT calculations (see Figure 12) show that both the $(PtCO)_2$ and the $Fe_3O_4(001)$ support must adopt a metastable configuration for the reaction to proceed at the temperature observed, because this ultimately reduces the energy required to stabilize the $Pt_2CO$ intermediate. This then breaks into a Pt atom and a PtCO, which can diffuse at elevated temperature, explaining why a mixture of adatoms and larger clusters remain in STM images after the reaction.

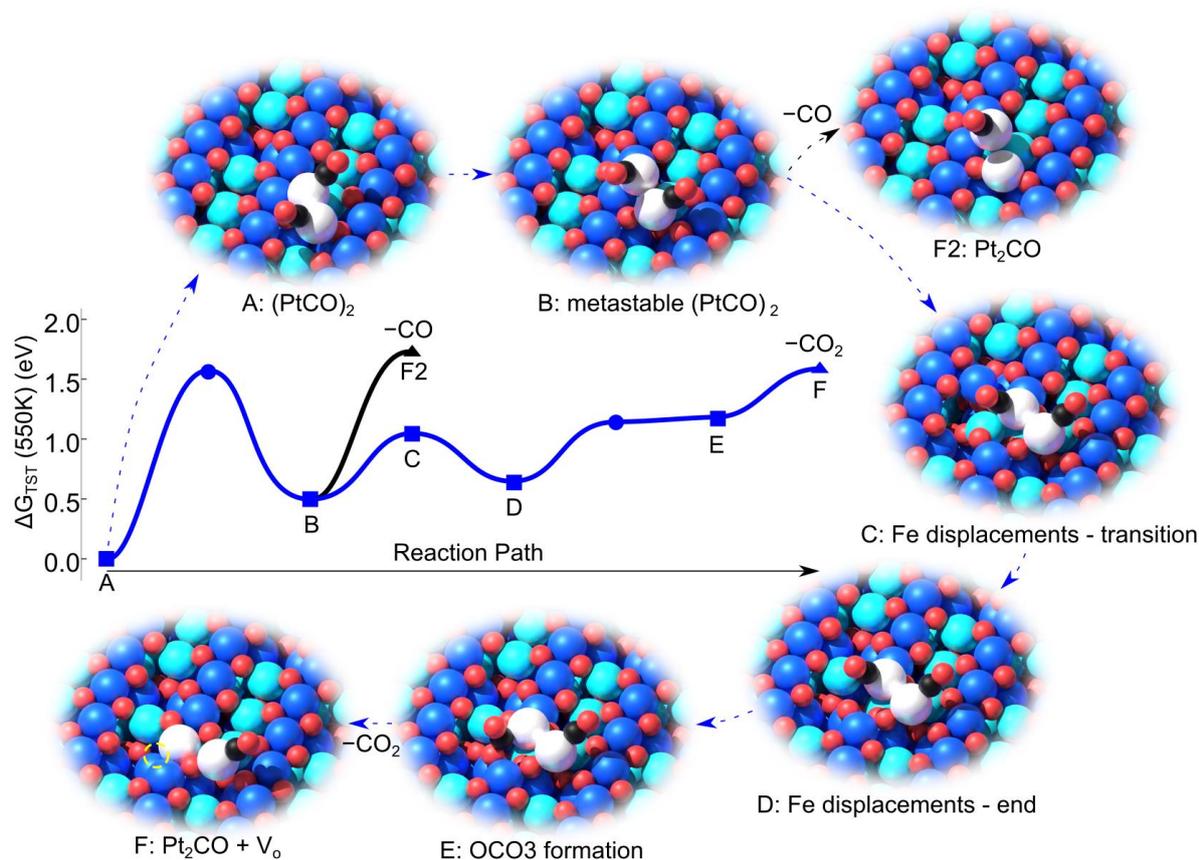

**Figure 12: Reaction scheme for the (PtCO)$_2$ species on Fe$_3$O$_4$(001).** The reaction occurs via a metastable configuration of the (PtCO)$_2$ and the Fe$_3$O$_4$(001) support, which allows extraction of lattice oxygen at minimum energetic cost. In the schematics, the Fe$_{oct}$ and Fe$_{tet}$ of the Fe$_3$O$_4$(001) support are dark blue and cyan, respectively. O atoms are red, Pt are white, and the C and O in CO are black and red, respectively. Figure reproduced from ref. [119]. Copyright 2022 American Association for the Advancement of Science under CC-BY license (https://creativecommons.org/licenses/by/4.0/)..

### 3.3.5 CO adsorption trends on Fe$_3$O$_4$(001)-based SACs

Recently, a systematic investigation of CO adsorption on Fe$_3$O$_4$(001)-based Au, Ag, Cu, Ni, Pt, Rh and Ir SACs was published.[112] The CO desorption temperature observed in TPD was converted to desorption energy assuming an ideal lattice gas and compared directly to the results of DFT calculations. The results reveal similar trends to those observed for close packed metal surfaces (see Figure 13), with some key differences. The noble metals bind weakest, and Rh and Ir strongest. In most cases the single atoms bind CO stronger than the close packed metal surface, with Ni the exception where a weaker interaction is observed. The differences were interpreted using the density of states extracted from DFT+$U$ calculations, and it was concluded that the proximity of the d-states to $E_\text{F}$ was a primary factor, as it is for metal surfaces. Adatoms in the 2-fold coordination geometry are all close to a 1+ oxidation state, so the electronic structure differs from a metal surface. In most cases, this results in a shift in the d-band center of mass to higher energies. However, this electronic effect is modulated by a couple of factors that play only a minor role for metal surfaces. For example, CO adsorption causes significant relaxations, which lower the adsorption energy from that expected on the basis of the d-band position alone. Also, there is the possibility, discussed above, that adsorption weakens the adatom-support interaction so much that a mobile metal carbonyl is created, which leads to sintering.

Overall, it was shown that the observed behavior can be understood in some cases by comparison to metal oxide surfaces, if similar oxygen coordination exists. In this scenario, CO competes with the oxide to bind the metal adatom, and strong CO binding destabilizes the atom on the surface. In the case of Rh and Ir, however, the preference for square planar and octahedral environments when CO adsorbs leads to a strengthening of the interaction with the support, and the twofold coordination of the adatom is easily modified by adsorption. Ultimately, the results clearly demonstrate that the adsorption properties of SAC systems are more closely related to coordination complexes than metal nanoparticles, which should influence the metals selected for a specific reaction.

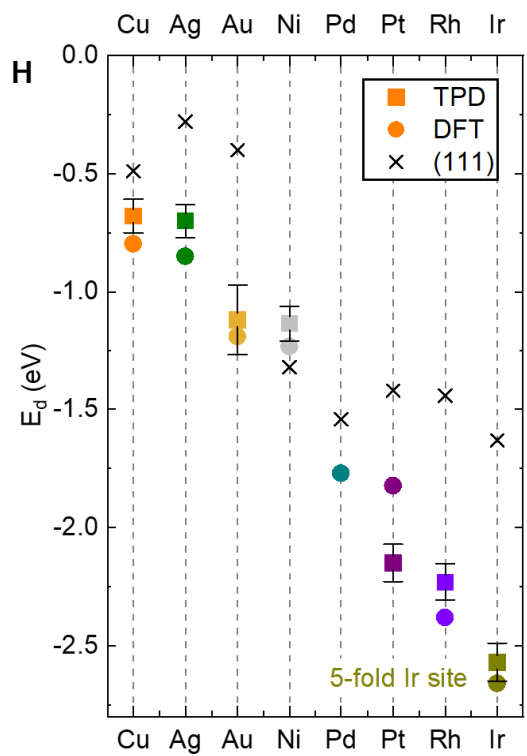

**Figure 13**: Plot of experimental and calculated CO adsorption/desorption energies, alongside experimental values for respective metal (111) surfaces. Error bars for the experimental data assume a temperature uncertainty of ±10 K (±20 K for Au). Figure adapted with permission from ref. [112]. Copyright 2021 American Association for the Advancement of Science.

### 3.3.6 $H_2$ activation trends on $Fe_3O_4(001)$-based SACs

Dohnalek and coworkers studied $H_2$ activation on the $Pd_1/Fe_3O_4(001)$ system using STM and DFT,[120] and found that a very high density of surface hydroxyls is created when $H_2$ is exposed to a surface with a low density of adatoms. This suggests that $H_2$ dissociation occurs, followed by spillover onto the oxide support to form surface OH groups. While OH diffusion is slow on the surface, it can be assisted by water, and this mechanism led to the migration of OH groups away from the Pd atoms. The experimental results for Pd matched well to the barriers predicted by DFT-based calculations, which gives confidence in the accuracy of the theory for this system. Similar calculations were then performed for a variety of metals (see Figure 14), and it was predicted that Pt will behave similarly to Pd, while heterolytic dissociation can also occur on Rh and Ir, leading to a possible equilibrium with hydride-hydroxyl pairs. For Ru, the hydride–hydroxyl pair becomes strongly preferred.

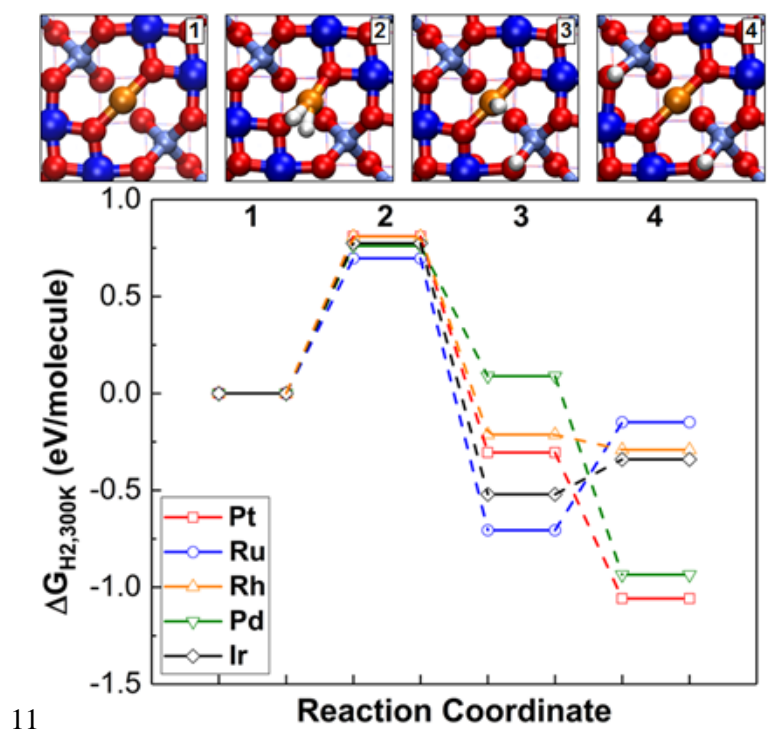

**Figure 14**: DFT-predicted energy pathway of $H_2$ dissociation mechanism on a single metal (Pd, Pt, Rh, Ir, and Ru) atom on $Fe_3O_4(001)$, corrected by gas-phase entropy at 300 K. In the schematics, oxygen atoms are red, Fe are blue, and the two-fold coordinated metal adatom is orange. H atoms are white. Reprinted with permission from ref. [120]. Copyright 2019 American Chemical Society.

## 3.4  Conclusions

Of the iron oxide surfaces, α-$Fe_2O_3$(0001) is most widely assumed in high surface area studies, but the adatom geometry generally assumed in DFT studies has not been reported by even a single experimental work. This is mainly because α-$Fe_2O_3$(0001) exhibits a variety of complex superstructures when prepared in UHV, and is thus extremely challenging to prepare and poorly suited as a model system. However, it is questionable how representative such UHV-specific terminations of hematite would be in realistic conditions. Experiments have shown hydroxylation of α-$Fe_2O_3$(0001) surfaces in the presence of even low background pressures (<$10^{-4}$ Torr) of water,[121] and theory also predicts high stability of hydroxylated surfaces.[122] Therefore, a more realistic approach to construct a model system for this termination may be to intentionally hydroxylate it, though this may be challenging in UHV.

At the other extreme, $Fe_3O_4$(001) has proved to be an excellent model system in that it stabilizes high loadings of single atoms at room temperature. This has facilitated fundamental studies of a range of SAC properties with good agreement between DFT calculations and experimental work, including adsorption trends for simple molecules on a range of different elements. Unfortunately, the structural modifications induced by water mean that it will be difficult to correlate local coordination to chemical reactivity at elevated pressures.

Finally, we have recently begun investigating the α-$Fe_2O_3$($1\bar{1}02$) surface as a SAC model system, and found it to be highly promising. This facet is much easier to prepare than the (0001), and single Rh adatoms were stabilized at room temperature by coadsorbed water.[104] Since the support in most powder catalyst works is hematite, rather than magnetite, we find this to currently be the most promising iron oxide model system for bridging high surface area studies and theory.

## 4. Ceria ($CeO_2$)

Cerium oxide is a clear example of a catalyst support that strongly participates in reactions, not only in terms of electronic effects modifying the catalytic properties of supported metals, but also acting as an oxygen reservoir. Due to the capability of Ce to reversibly transform between the two stable oxidation states of $Ce^{4+}$ and $Ce^{3+}$, ceria can easily exchange oxygen with the environment. This has multiple implications for single-atom catalysis: An abundance of different types of defects related to oxygen vacancies in the surface, subsurface or at steps supplies potential sites to stabilize adatoms at low coverage. However, the low barriers for creating and repairing these defects also implies that they might change significantly under reaction conditions, potentially destabilizing the adatoms. Facile oxygen vacancy creation can also be relevant if oxygen for the reaction is supplied directly from the surface, i.e. in a Mars-van Krevelen mechanism.

The typical defects on $CeO_2$(111) have been studied extensively. Surface and subsurface oxygen vacancies can easily be introduced by annealing in reducing conditions, and give rise to $Ce^{3+}$ ions at the surface.[123] For many metals, the most relevant adsorption sites appear to be at step edges,[124] which have also been studied in detail.[125,126]

Adsorption studies on a wide variety of metals on stoichiometric or reduced $CeO_2$(111) can be found in the literature, though mostly at high coverages. Generally, these can be divided into metals that become fully oxidized and form mixed oxides with ceria at room temperature, such as Al, Ga and Sn,[127-129] and metals that quickly form metallic clusters. With few exceptions, clusters tend to preferentially decorate step edges on stoichiometric ceria,[130-132] or nucleate at surface defect sites on oxygen-deficient ceria films.[133,134] It is worth noting that whether charge transfer occurs, and whether the ad-metals are reduced or oxidized, appears to be highly dependent on both the metal and the oxidation state of the ceria film.[135-139]

## 4.1  Pt, Pd, Ni on $CeO_2$

Stabilisation of single Pt adatoms on $CeO_2$ was studied extensively by the groups of Matolín and Libuda. A detailed review of the Pt/$CeO_2$ system already exists,[140] but we will summarize the findings here to put them in the context of SAC. While the ideal $CeO_2$(111) surface does not provide sites to trap single Pt atoms, Matolín and Libuda used SRPES and DFT to identify a "nanopocket" stabilizing $Pt^{2+}$ in a square-planar coordination on {100} nanofacets, with an exceptionally high adsorption energy of −678 kJ/mol.[141] TEM shows that such nanofacets indeed exist on $CeO_2$ particles prepared by magnetron sputtering. In the surface science studies, the nanofacet sites were prepared by co-depositing Ce and Pt in an oxygen atmosphere on a $CeO_2$(111) film, under conditions suitable for forming $CeO_2$ nanoparticles. While these nanoparticles on the $CeO_2$(111) surface can still be imaged by STM, the roughness of the surface does not allow direct confirmation of the Pt adsorption site by scanning probe techniques. However, the $Pt^{2+}$ particles are shown to resist reduction, sintering, and bulk diffusion up to the highest temperatures that could be achieved on that surface (ca. 750 K), in agreement with the high stability predicted by DFT. Interestingly, the calculated adsorption energy of Pt in the nanopocket exceeds the cohesive energy of bulk Pt (−564 kJ/mol), and DFT predicts that abstracting Pt from metallic clusters is possible (Figure 15).[141] Indeed, at least partial redispersion from metallic particles to $Pt^{2+}$ was later shown in experiment.[140]

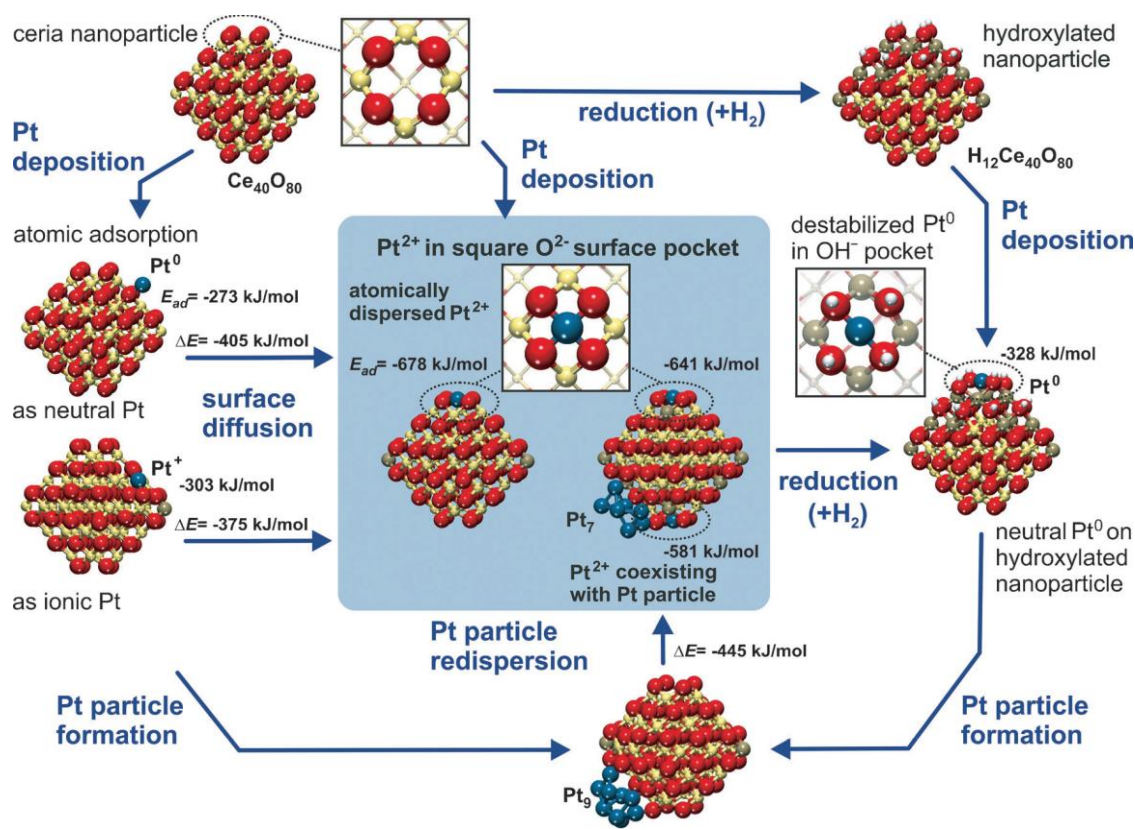

**Figure 15:** Structure and energetics of the anchored $Pt^{2+}$ species on ceria nanoparticles determined by theory. The $Pt^{2+}$ is strongly bound at the {100} nanofacets of the ceria nanoparticle. Color coding of atoms: red O, beige $Ce^{4+}$, brown $Ce^{3+}$, blue Pt, white H. Reproduced with permission from ref. [141]. Copyright 2014 John Wiley and Sons.

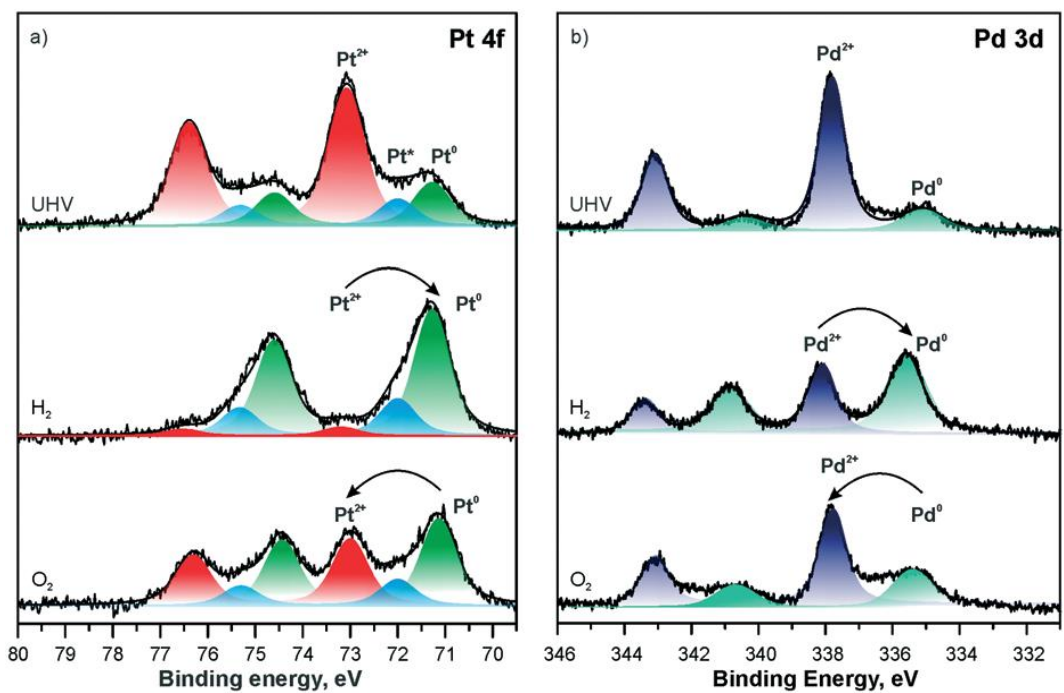

**Figure 16:** The development of Pt 4f (a) and Pd 3d (b) spectra obtained from Pt–CeO$_2$ and Pd–CeO$_2$ films, respectively, following the annealing in UHV (top), under H$_2$ (middle), and O$_2$ (bottom) atmosphere. Reproduced from ref. [140] with permission. Copyright 2017 Royal Society of Chemistry.

The high stability of Pt$^{2+}$ in a surface site seems to recommend this system as a single atom catalyst. However, the Pt atoms appear to be rather inactive: No CO is adsorbed above 110 K,[142] and no dissociation of molecular hydrogen is observed.[143] This is not surprising, as the square-planar "surface pocket" already provides an ideal environment for the Pt$^{2+}$ ion in d$^8$ configuration, which disfavours further bonds. However, the catalysts can be activated by reducing Ce$^{4+}$ to Ce$^{3+}$, which destabilizes the single-atom sites (Figure 16). Annealing in hydrogen (when some metallic Pt is present to initialize hydrogen dissociation),[143] in methanol,[144] or depositing Sn as a reducing agent[145] thus accumulates the Pt$^{2+}$ single atoms to sub-nanometre clusters, which serve as a "working state" for reaction. The authors speculate that a Pt-CeO$_x$ catalyst with ideal Pt loading will be able to reversibly cycle between atomically dispersed Pt$^{2+}$ species and the active sub-nanometre particles.[141] In this scenario, the single-atom sites are not involved in SAC in a literal sense, but are still highly relevant for the long-term stability of the catalyst, as the regular re-dispersion prevents accumulation to larger clusters in the long run.

A structural motif similar to the "nanopockets" was identified by Matolín's group at monolayer-high step edges on $CeO_2(111)$.[124] By adjusting the step density and the density of surface oxygen vacancies (Figure 17), they showed with STM and PES that high step density allows accommodation of large amounts (0.05 ML) of $Pt^{2+}$ without any cluster formation. In contrast, surface oxygen vacancies only serve as anchoring points for metallic clusters, but do not prevent their formation. While the Pt atoms are again not visible in STM, the authors identify likely adsorption sites at steps by DFT. These sites again feature square-planar $PtO_4$ moieties and very high adsorption energies (5.0-6.7 eV). There is both theoretical[146] and experimental[147] evidence that stable surface peroxo units form after exposure of $CeO_2(111)$ to molecular oxygen, and that accommodating excess O atoms at steps allows to increase the amount of $Pt^{2+}$ species established after the deposition of Pt, thus maximising the "single-atom" capacity of the surface. Very recently, cycling between single Pt atoms and Pt clusters has also been demonstrated experimentally for Pt anchored at step sites.[148] It should be noted that since steps are present on any $CeO_2(111)$ film, the step-supported type of $Pt^{2+}$ species identified by Matolín would almost certainly also be present on the $CeO_2(111)$ films supporting $CeO_2$ nanoparticles discussed above. Even if no step-like sites exist on the nanoparticles directly, they should still be abundant on the underlying film. This implies that the results on $Pt^{2+}$ in "nanopockets", showing their high stability but also inactivity,[141-143] are transferrable to Pt supported at monolayer steps.

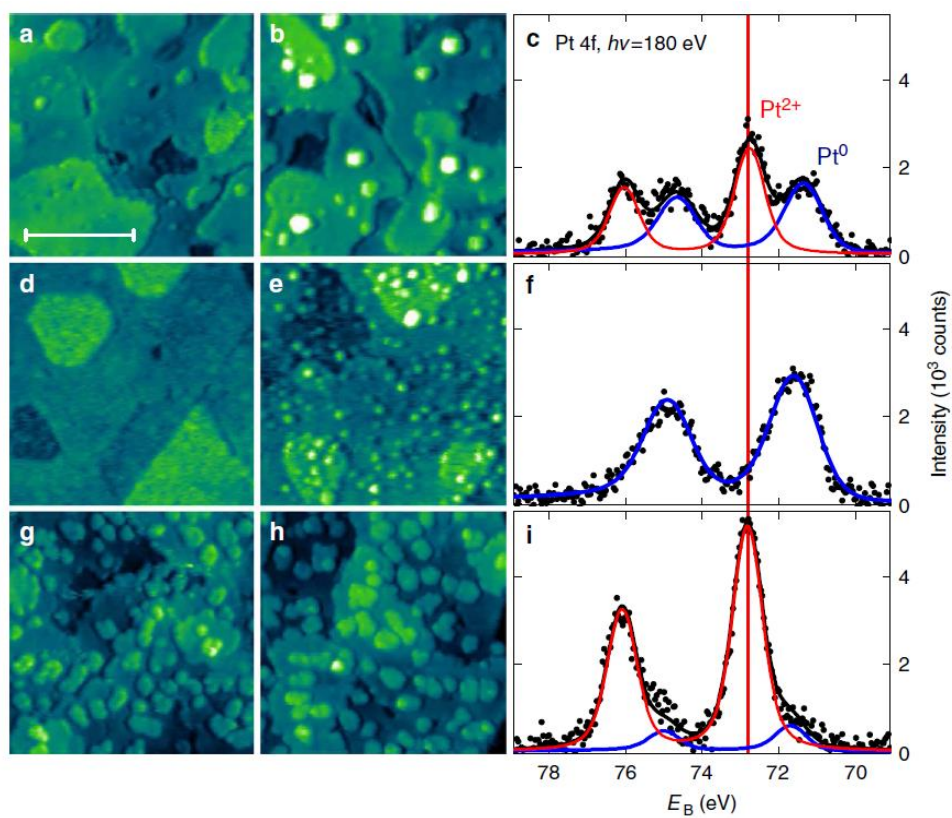

**Figure 17:** Nucleation of Pt and stabilization of $Pt^{2+}$ on ceria surfaces containing controlled amount of surface defects. (a–c) $CeO_2$(111) surface with low density of surface oxygen vacancies and ML-high steps. (d–f) $CeO_{1.7}$ surface with increased density of surface oxygen vacancies. (g–i) $CeO_2$(111) surface with increased density of ML-high steps. (a,d,g) STM images of clean surfaces before deposition of Pt. (b,e,h) STM images after deposition of 0.06 ML Pt and annealing at 700 K in UHV. All STM images 45×45nm², tunnelling current 25-75 pA, sample bias voltage 2.5-3.5 V. Scale bar, 20nm (a). (c,f,i) PES spectra of the Pt deposit after annealing. All PES spectra were acquired with photon energy hν=180 eV (black points). Fits indicate metallic ($Pt^0$, blue line) and ionic ($Pt^{2+}$, red line) contributions to Pt 4f signal. $E_B$ is the photoelectron binding energy. Reproduced with permission from ref. [124]. Copyright 2016 Springer Nature under CC-BY license (https://creativecommons.org/licenses/by/4.0/).

Even more generally, many nanoparticle studies of $Pt/CeO_2$ report Pt to be situated on facets other than [111], which have not been as thoroughly investigated by surface science methods. However, a recent study of Pt on a realistic ceria support, combining DFT with x-ray absorption spectroscopy (XAS), IRAS and XPS,[149] suggests that the results of Libuda and Matolín are also generalizable to other facets. While this study finds higher stability for Pt on the [110] and [100] facets, the stable site is always a four-fold coordinated pocket, and single-site Pt is initially inactive towards CO, $C_3H_6$ and $CH_4$ oxidation. Catalytic activity again only sets in once Pt is significantly reduced and sinters to small clusters.

Generalizing these results from Pt, a DFT investigation of adsorption energies at the {100} nanofacet site found that, for 11 different metals, the nanofacet site was always preferred to adsorption on a metal nanoparticle.[150] The largest differences in adsorption energies were found for group X metals (Ni, Pd, and Pt) and for Fe, Co, and Os, indicating that these metals should be stabilized against sintering by the "nanopocket". Investigations on Pd and Ni indeed show that these metals are likewise stabilized in a 2+ state.[151] However, unlike Pt, both Pd and Ni segregate to their native oxides under some conditions, and both can be stabilized in ceria bulk sites, leading to bulk diffusion above 600 K. Furthermore, annealing in hydrogen did not lead to a change of oxidation state for Ni, suggesting that the "active state" of metallic particles is inaccessible, likely because the $Ni^{2+}$ species is too stable.

While the experimental results discussed above suggest that single Pt atoms on ceria are only activated by cluster formation, a more recent theoretical study has instead proposed a reaction pathway in which single Pt atoms on $CeO_2$(100) can be both stable and active after a reductive activation step in sufficiently high pressures of $H_2$ or CO, which puts them in a transient two-fold coordinated state.[152] Crucially, the reaction pathway involves phonon-assisted switching of the platinum charge state during reaction through electron injection to (and recovery from) the support. While these transient configurations and charge states may be difficult to characterize experimentally, the authors argue that dynamic charge transfer needs to be taken into account in modelling, rather than assuming a fixed Pt charge state.[152] Such a pathway could in principle be reconciled with the studies showing clustering if active single Pt atoms exist in a preparation window in which ceria is sufficiently reduced, but at low enough temperature to prevent agglomeration.

Very recently, Wan and coworkers have explored the effect of pre-treating CeO$_2$(111) films with oxygen plasma. They conclude that the plasma treatment causes nanostructuring of the surface, as well as formation of peroxo species in the surface, as shown schematically in Figure 18. Even relatively high loadings of 0.2 ML Pt are apparently stabilized as single atoms on surfaces prepared in this way. Crucially, while at least some of the Pt atoms are in inactive "nanopocket" sites, other adsorption geometries are also present. The Pt single atoms are thermally stable and active for CO oxidation, as the authors demonstrate both for the thin film model system and for powder catalysts.

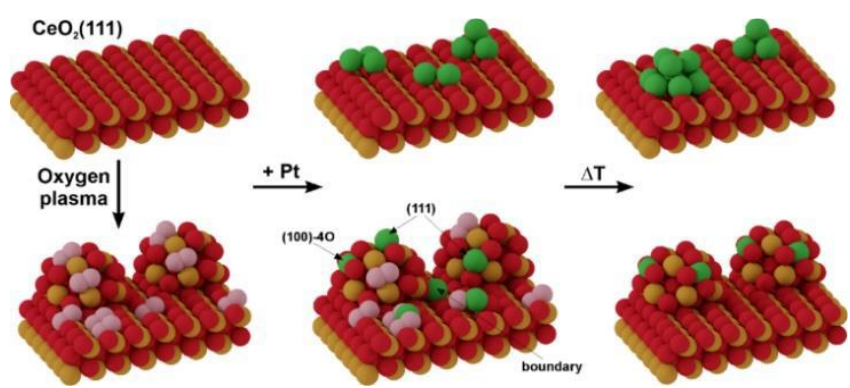

**Figure 18**: Schematic representation of the interaction of Pt atoms with pristine and oxygen plasma-treated CeO$_2$(111) films. Upon deposition on a stoichiometric CeO$_2$(111) surface, Pt forms small clusters which aggregate into larger Pt nanoparticles at elevated temperatures. Plasma pre-treatment of the CeO$_2$ surface produces peroxo species and induces surface restructuring, resulting in small ceria nanoparticles, which act as anchoring sites either directly upon Pt adsorption or through surface migration of peroxo-stabilized Pt single atoms. Color code: Ce - gold, O - red, Pt - green, and peroxo O$_2^{2-}$ - pink. Reproduced with permission from ref. [153]. Copyright 2022 Wiley-VCH GmbH under CC-BY license (https://creativecommons.org/licenses/by/4.0/).

## 4.2. Au, Ag, Cu on CeO$_2$

While gold supported on ceria has been shown to be a promising single-atom catalyst,[154] its coordination to the ceria support in the active state is contentious. On the stoichiometric $CeO_2$(111) surface, theory predicts adsorption on oxygen bridge sites to be slightly preferred over top sites.[155] However, it has been pointed out that the preferred site and the oxidation state of the gold atom on ceria strongly depends on the chosen theoretical method.[156] Experimentally, when gold is deposited on stoichiometric $CeO_2$(111) at 10 K, it is found in both oxygen top and oxygen bridge sites.[157] Since gold accumulates at step edges at room temperature,[130] this is likely due to limited mobility at 10 K. On the terrace sites, accumulating gold forms upright dimers and compact 3D clusters. From this and the absence of characteristic fingerprints of charged species in STM, the authors infer a close-to-neutral charge state of the aggregates.[157]

On oxygen-deficient $CeO_2$(111), surface and subsurface oxygen vacancies are present, introducing $Ce^{3+}$ sites at the surface.[123] When gold was dosed on this surface at 10 K, some Au got trapped in surface oxygen vacancy sites, from which it could not be removed with the STM tip anymore.[158] However, a later report[139] points out that these defect sites are only rarely occupied. Rather, gold mostly binds to bridge sites when deposited at 15 K, then clusters at step edges upon annealing to 400 K, while undecorated surface oxygen vacancies remain visible (Figure 19). The authors conclude that while the oxygen vacancy is thermodynamically favoured, diffusion into the vacancy site is kinetically hindered up to above 395 K. According to their DFT calculations, the large diffusion barrier (≈1.0 eV) is a consequence of Au having to change its charge state from +1 on the pristine surface, to 0 near the vacancy, to −1 in the vacancy, in order to diffuse there.[139]

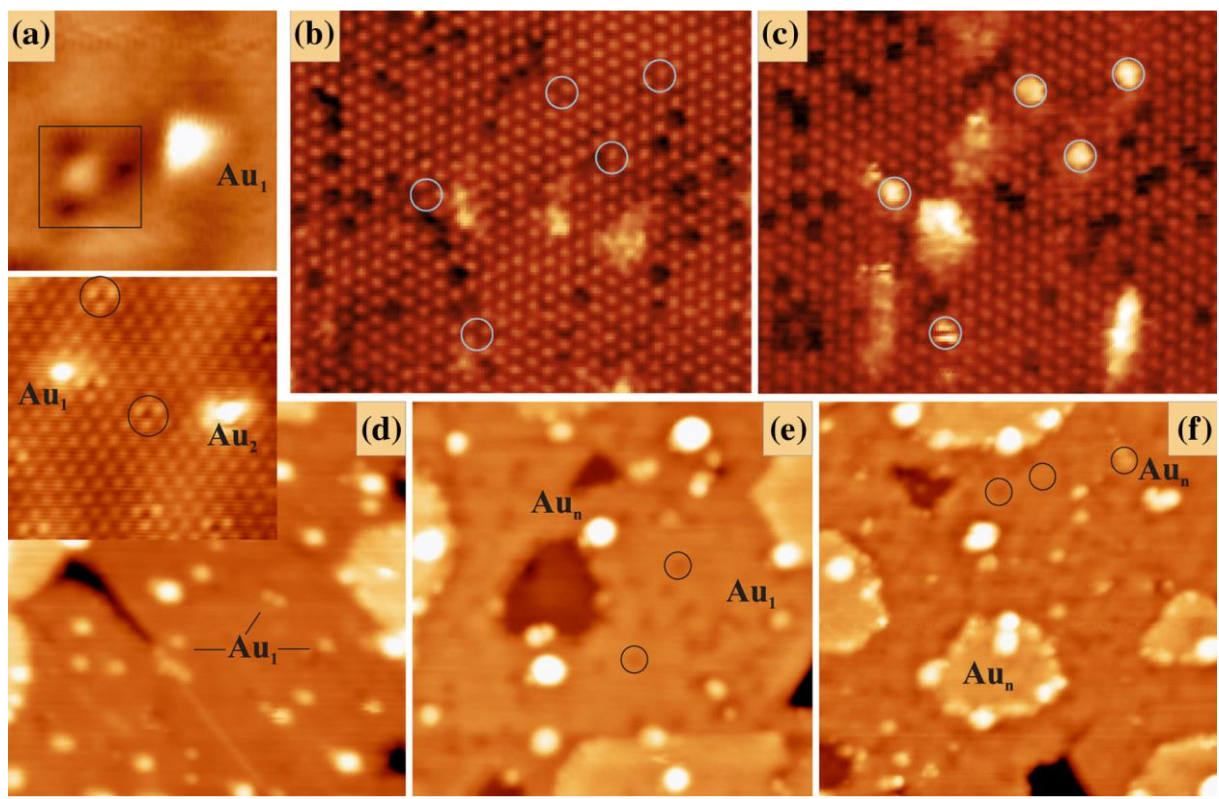

**Figure 19:** (a) High-resolution STM image of a regular and a defect-bound Au atom (box) deposited on $CeO_{2-x}$/Ru(0001) at 15 K (+2.5 V, 3.8 × 3.0 nm$^2$). (b),(c) Identical surface region on a √3-reconstructed $CeO_{2-x}$/Pt(111) film before and after exposure of 0.05 ML Au at 15 K (−3.0 V, 10.5 × 8.0 nm$^2$). Adatoms and their binding sites on the pristine surface are marked by circles. (d) $CeO_{2-x}$/Ru(0001) after gold deposition at 15 K and (e) after annealing to 200 and (f) 400 K (2.5 V, 30 × 30 nm$^2$). The inset in (d) shows a close-up of the main image; some $V_O^S$ defects are encircled. Reprinted figure with permission from ref. [139]. Copyright 2016 by the American Physical Society.

Interestingly, gold was found to not only interact with the surface oxygen vacancies directly, but also with the $Ce^{3+}$ ions introduced by subsurface oxygen vacancies.[158] Dosed onto the defective surface at 10 K, Au atoms frequently appeared in pairs at a mean distance of 7.6 Å (two surface lattice constants), which DFT identifies as the expected spacing of the $Ce^{3+}$ ions. The appearance of the paired features and the DFT results further suggest that charge transfer occurs at these sites, creating $Au^-$ species.

Ag and Cu single atoms on $CeO_2(111)$ have been explored much less than Au, but they appear to follow similar trends. Both accumulate mainly at step edges, with Ag interacting more strongly with the reduced surface,[132,159,160] as is the case for Au. The experimental and theoretical evidence suggests that for both Ag[132] and Cu,[137] single atoms and small clusters are oxidized on the pristine $CeO_2(111)$ surface, but adopt a negative charge state when they are localized at surface oxygen vacancies. Interestingly, calorimetric studies show that while Ag interacts more strongly with reduced films than with the stoichiometric surface,[133] the opposite is true for Cu,[161] indicating different interaction with surface oxygen vacancies or with $Ce^{3+}$ sites.

In the model studies discussed so far, the potential host sites considered for the catalyst metal are usually oxygen vacancies or ad-sites. In contrast, in nanoparticle studies, the active sites are often assigned as cerium substitution sites, based on TEM.[154,162] Such configurations have been considered for Cu and Ag in two recent works,[163,164] although the applied doping levels (ca. 10 at. %) were much higher than in SAC. Experimentally, both films were found to be more reducible by annealing in UHV than pure $CeO_2$. This is in agreement with DFT, which predicts spontaneous formation of surface oxygen vacancies near the modifier cations.[163] However, the Ag-modified films showed a lower concentration of $Ce^{3+}$ cations than pristine ceria even in the presence of more oxygen vacancies.[164] This is explained by reduction of $Ag^{2+}$ to $Ag^+$ being more favourable than reduction of $Ce^{4+}$. This again highlights the difficulty in assigning sites based only on TEM in combination with DFT.

Short-lived single Au atoms have also been proposed as the active species in CO oxidation over ceria-supported Au nanoparticles based on molecular dynamics simulations.[165] In the proposed mechanism, adsorbed CO induces gold atoms to break away from a nanoparticle as $Au^+$-CO and diffuse on the surface, occupying on-top positions on surface oxygen atoms. Such a mechanism would probably not be recognized as a "single-atom" catalyst in most experiments, and indeed it is debatable whether it should be considered as such, as the dynamic $Au^+$-CO complex can likely only be created in the presence of nanoparticles. Nonetheless, like the dependence of single atom stability on the ceria oxidation state, this again highlights the strong influence of the environment on ceria-supported metals.

## 4.3 Rh on CeO$_2$

The case of Rhodium is interesting as a PES study from 2016 has shown more clearly than for other metals that it has the capacity to either oxidize or reduce a CeO$_2$(111) film, depending on the film's initial stoichiometry (Figure 20).[138] While there are no atomic-scale images, the SRPES data in combination with DFT strongly suggest that Rh can either be in a cationic state on the stoichiometric surface, or become anionic when occupying an oxygen vacancy. It is interesting to note, however, that in an STM study of larger amounts of Rh deposited on CeO$_2$(111) at room temperature, clusters preferentially decorated steps, with no difference between stoichiometric and reduced films.[131] This may suggest that in contrast to what has been described for gold,[134] Rh ions at oxygen vacancy sites do not act as nucleation centres for cluster formation.

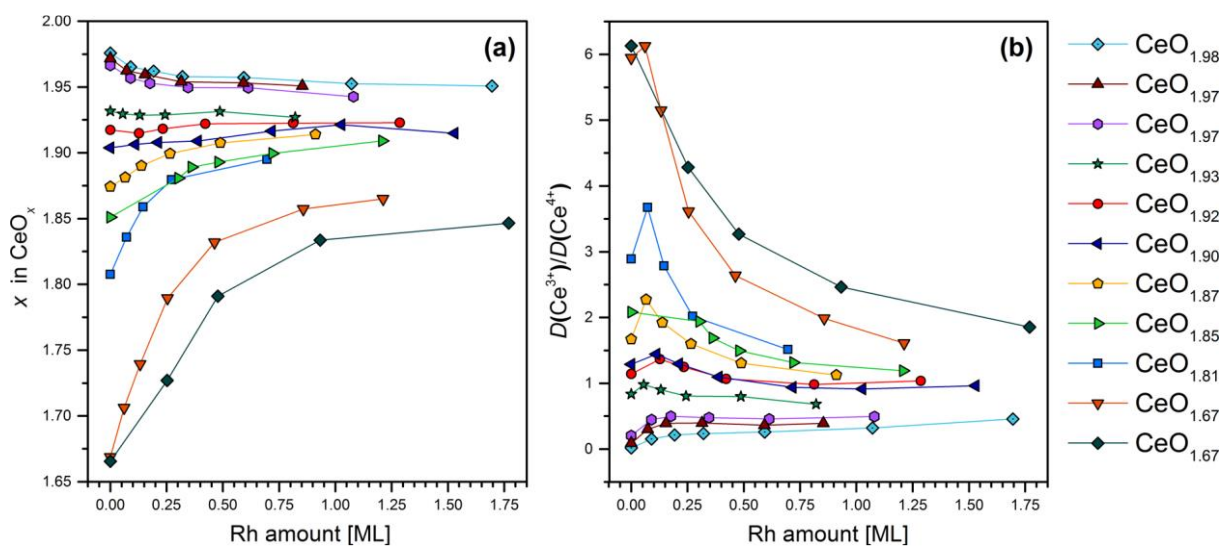

**Figure 20:** Evolution of the degree of CeO$_x$ reduction for various cerium oxide stoichiometries (1.98 > x > 1.67) during the consequent depositions of rhodium estimated from (a) XPS and (b) SRPES measurements. Reprinted with permission from ref. [138]. Copyright 2016 American Chemical Society.

## 4.4  Conclusions

On the one hand, ceria is an excellent SAC model system in that a variety of elements can be stabilized as single atoms at steps or nanofacets when the surface is sufficiently oxidized. On the other hand, these stable sites appear to be universally inactive, and activity is obtained only when the single atoms are destabilized and sinter to small clusters.[140] Recent work showing Ag and Cu in Ce substitutional sites seems promising,[163,164] but it remains to be seen whether these dopant atoms remain at the surface and are accessible to adsorbates. Single atoms adsorbed on $CeO_2(111)$ have been studied at low temperatures, and especially the interrelation of charge state and adatom diffusion, which prevents Au adatoms from reaching the energetically favourable $V_O$ sites,[139] is an interesting fundamental result. However, these adatoms generally sinter at room temperature, and so are probably not the active site observed in high surface area studies. To our knowledge, there has not been any experimental work investigating stabilization or destabilization of adatoms by co-adsorbates.

## 5. Magnesium Oxide (MgO)

MgO has received extensive attention over the years both as a catalyst support and as a thin insulating oxide. In fact, MgO was the basis for arguably one of the earliest works demonstrating single atom catalysis on a model system, namely acetylene cyclotrimerization on $Pd_1$ at 300 K.[166] Several types of defects which may play a role in stabilizing single adatoms and small clusters have been identified. Oxygen vacancies are found mainly at steps and can appear in the form of $F^0$, $F^+$ or $F^{2+}$ colour centres, where $F^+$ and $F^0$ correspond to one and two electrons trapped in the vacancy site, respectively. It has been demonstrated that existing vacancy sites can be charged by electron irradiation or simply by scanning with an STM tip,[167-169] and new defect sites are created by higher electron doses (Figure 21).[168]

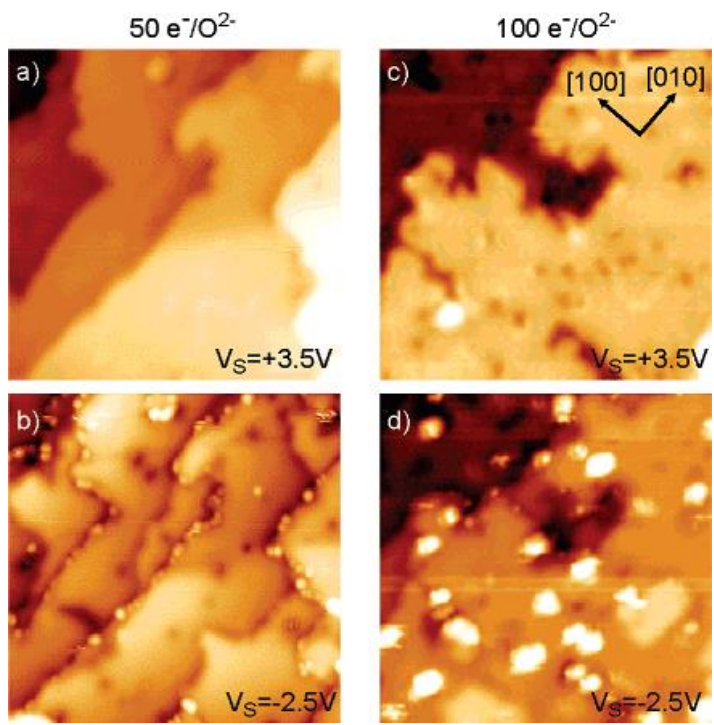

**Figure 21**: STM images (30 × 30 nm) of a 4 ML MgO(001) film grown on Ag(001) after bombardment with low (a, b) and high (c, d) electron doses. (a) and (c) were obtained at a sample bias of $V_S = +3.5$ V, and (b) and (d) at $V_S = -2.5$ V. Reprinted with permission from ref. [168]. Copyright 2006 American Chemical Society.

A second family of defects are the so-called $(H^+)(e^-)$ colour centres, where deep electron traps are created in the vicinity of adsorbed protons. These defects also occur mainly at edges and kinks, and can be created either by first adsorbing hydrogen, splitting it heterolytically and then oxidizing the hydride by UV irradiation, or by first creating an $O^-$ centre through UV irradiation and then splitting $H_2$ there to obtain $H^+$ and a hydrogen radical.[170-172] It is worth noting that while the existence of these $(H^+)(e^-)$ centres has been demonstrated clearly by a combination of theory and electron paramagnetic resonance (EPR) spectroscopy, they have to the best of our knowledge not been identified by scanning probe techniques.

Finally, a defect predicted by theory to be relevant for trapping adatoms is the neutral divacancy, in which an entire MgO unit is removed at the surface.[173,174] Experimentally, evidence for this defect is scarce, though this may in part be explained by the fact that it may not be visible with most spectroscopic techniques (including EPR). Non-contact AFM images of cleaved MgO do indeed show pits that may correspond to divacancies, but unambiguous evidence is missing.[175,176]

## 5.1. Au on MgO

Probably the most extensively studied ad-metal on MgO is gold. An early experimental work on size-selected clusters demonstrated that charge (~ 0.5 e) is transferred into adsorbed gold clusters and single adatoms on defect-rich films.[177] This was tentatively attributed to charge transfer from $F^+$ centres. While single atoms were essentially inert, $Au_8$ was shown to be active for CO oxidation on defect-rich films, where charge transfer occurs, but not on defect-poor films, where it does not.[177] In this, gold behaves differently from e.g. $Pd_8$, which was shown to be active no matter the defect concentration of the film.[178] The negative charge of Au in this work is contrasted by more recent findings by the Freund group (Figure 22),[179] who showed initial formation of positively charged gold, which they – also tentatively – attributed to charge transfer from gold into existing deep electron traps on the surface, such as grain boundaries of the MgO film. While these results seem contradictory at first glance, it seems entirely possible that the exact interaction of adatoms with a given MgO film depends on the type and density of defects, as well as on the initial charge state of these defects.

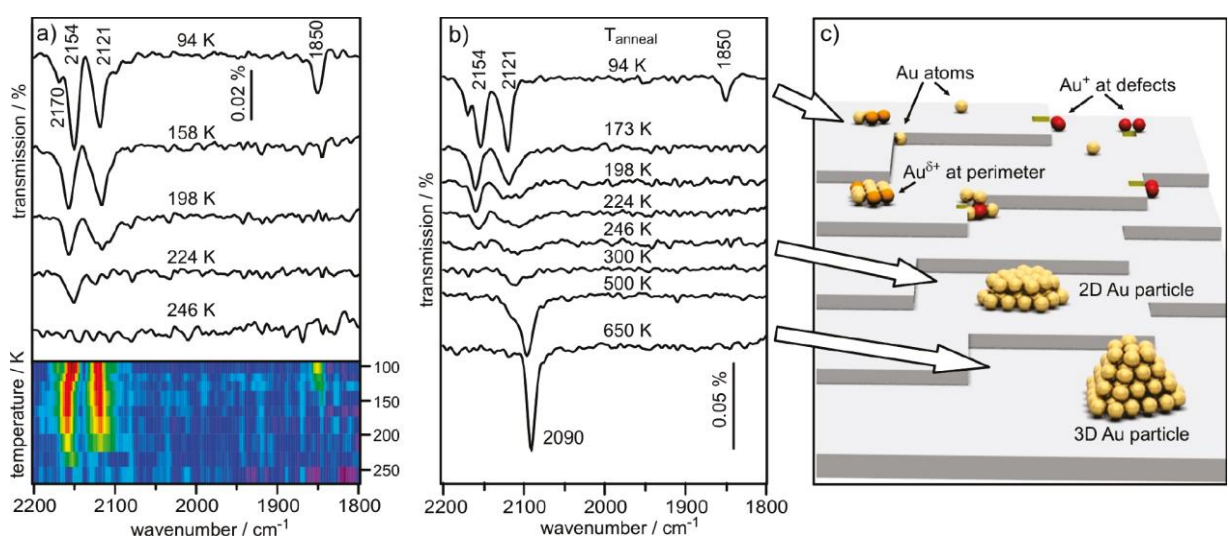

**Figure 22:** (a) IR spectra of CO adsorbed on 0.02 ML Au/13 ML MgO(001)/Ag(001) as a function of temperature. The spectra were collected at the indicated temperature. The lower panel presents the results as an image plot with red being intense and blue representing no absorption. (b) IR spectra of CO adsorbed on 0.02 ML Au/13 ML MgO(001)/Ag(001) as a function of annealing temperature. The spectra were collected after recooling to 90 K and dosing with CO. (c) Model of the Au/MgO(001) surface representing the nature of Au species formed at various annealing temperatures as deduced from the IR spectra shown in part b. Reprinted with permission from ref. [179]. Copyright 2011 American Chemical Society.

Concerning nucleation sites for small clusters, theory predicts strong trapping of Au adatoms in oxygen vacancies and divacancies.[180] This seems to be confirmed by a low-temperature STM study in which a thin MgO film was first irradiated by electrons to introduce colour centres, and Au was then deposited at 5-8 K.[181] At these lowest temperatures, Au atoms initially adsorb as single atoms and dimers at terrace sites (Figure 23). After annealing to 30 K, the EPR signal corresponding to colour centres is quenched, suggesting that small gold clusters have formed there.

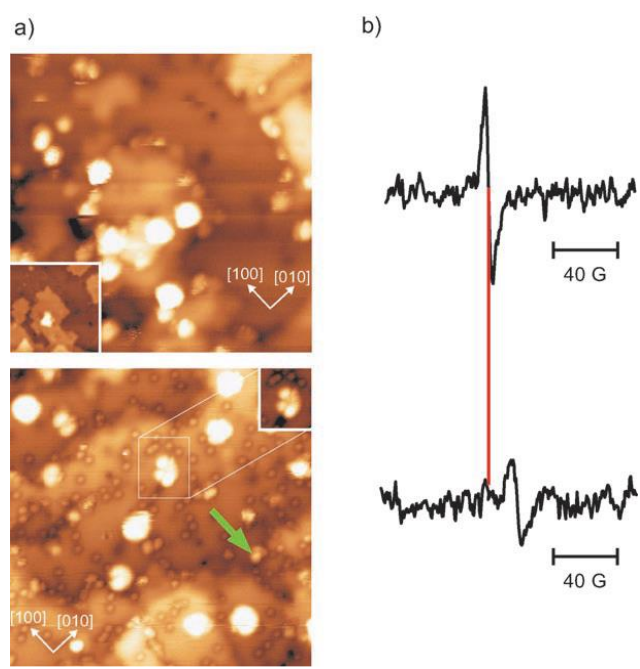

**Figure 23:** a) STM images (30 × 30 nm$^2$). Top: 3-4 ML MgO(001)/Ag(001) taken after electron bombardment, $V_s = -3.0$ V, $I_t = 8$ pA; inset: same area $V_s = 3.5$ V, $I_t = 9$ pA; bottom: the same preparation after deposition of 0.035 ML Au at 5-8 K; a dimer is indicated by a green arrow; inset:

nucleation on a color center, with adjusted contrast; $V_s$ = 1.3 V, $I_t$ = 10 pA; 30 × 30 nm². b) EPR spectra around g = 2, top: MgO(001) film on Mo(001) after low-dose electron bombardment; bottom: the same preparation after deposition of 0.015 ML Au at 30 K. The red line indicates the position of the color-center signal in both spectra. Reproduced with permission from ref. [181]. Copyright 2006 John Wiley and Sons.

Apart from the defects present, the properties of gold adsorption on MgO thin films also depend strongly on the thickness of the film. Theory predicts partial charge transfer from the metal substrate to gold for ultrathin MgO films on Ag or Mo, and similar binding energies of Au on O and Mg terrace sites.[182,183] Again, this is in good agreement with low-temperature STM results, which find Au exclusively adsorbing on O ions on 8 ML MgO films, but about equal occupancies of O and Mg sites on 3 ML films (Figure 24).[184] The underlying metal may contribute even more strongly if cations diffuse into the film and act as dopants. For example, MgO films grown on Mo(100) have been shown to contain Mo(V) centres, which appear to be situated at the surface.[185,186] It seems likely that this would also influence the properties of adatoms on the MgO film, as Mo dopants in CaO(001) films, for example, have been shown to donate charge to adsorbed Au clusters, changing their shape.[187]

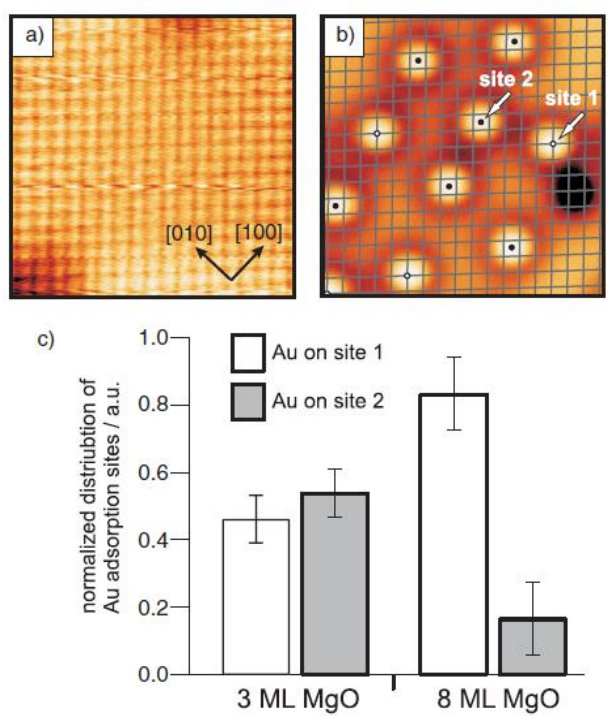

**Figure 24:** (a) Atomically resolved STM image (5 nm × 5 nm) on 3 ML MgO/Ag(001), $V_S$ = −0.02 V, $I_T$ = 7 nA; only one ionic sublattice is resolved. (b) STM image (5 nm × 5 nm) of Au atoms deposited at 5-10 K on 3 ML MgO/Ag(001), $V_S$ = −0.5 V, $I_T$ = 10 pA; the ionic sublattice extracted from (a) is superimposed revealing the different adsorption sites. (c) Distribution of adsorption sites for Au on 3 ML and 8 ML thin MgO films, respectively. Reproduced with permission from ref. [184]. Copyright 2007 American Physical Society.

## 5.2. Pd on MgO

Apart from Au, the ad-metal that has been explored in most detail (at low coverages) is Pd. Interestingly, an AFM study on Pd adsorption on MgO over a wide temperature range concludes that nucleation kinetics are governed by point defects, most of which are found on MgO terraces.[188] This is an interesting contrast to what one would expect for the case of Au: If Au mostly nucleates at colour centres,[181] and colour centres are found mostly at steps,[168] this would suggest clusters should predominantly decorate step edges. Therefore, it seems that Pd either shows quite different nucleation behaviour from Au, or the role of other defects (e.g. uncharged divacancies at terraces) in cluster nucleation is generally underestimated. Divacancies and $F^+$ centres have been identified as the most plausible nucleation sites by a theoretical study,[174] which also predicts steps to be poor trapping sites, but it should be noted that F centres at steps were not taken into account in this comparison.

As mentioned above, catalytic activity for single Pd atoms on MgO has been demonstrated already 20 years ago by TPD and IRAS.[166] Size-selected clusters were deposited on MgO at 90 K, then saturated with $C_2H_2$. Upon heating, single Pd atoms catalysed reaction to benzene at 300 K (Figure 25). This was attributed to charge transfer from defects into the Pd adatoms. Unfortunately, to our knowledge, the exact binding site of these Pd atoms has never been determined.

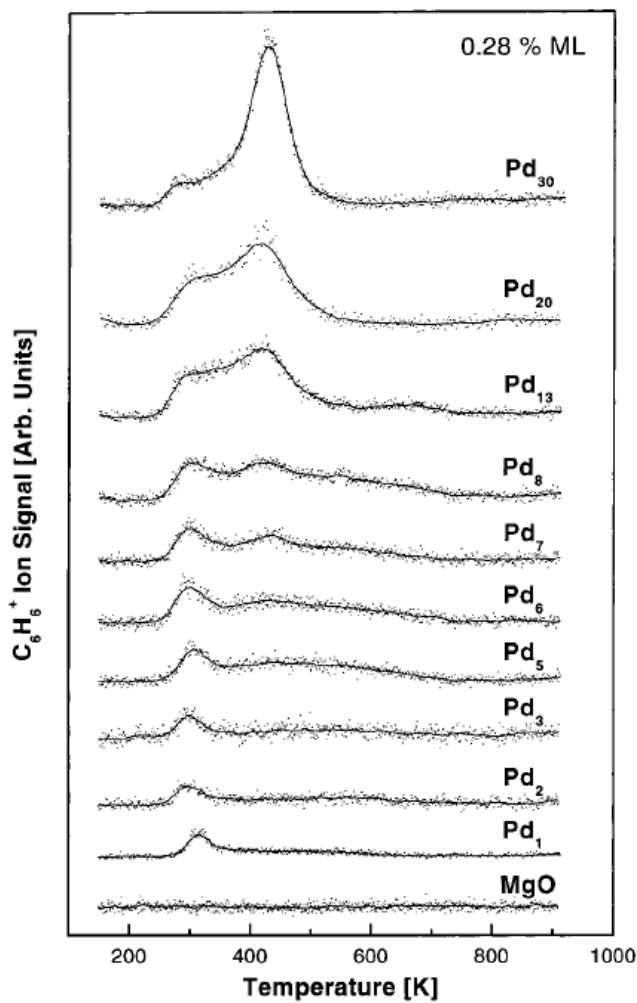

**Figure 25:** Catalytic $C_6H_6$ formation for different Pd cluster sizes obtained from temperature-programmed reaction experiments. The bottom spectrum shows that for clean MgO(100) films no benzene is formed. Dots, data; full line, data smoothing with adjacent averaging (25 points). Cluster coverage is 0.28% of a monolayer for all cluster sizes, where one monolayer corresponds to $2.25 \times 10^{15}$ atoms/cm$^2$. Reprinted with permission from ref. [166]. Copyright 2000 American Chemical Society.

## 5.3. Conclusions

MgO is clearly a challenging system to investigate with surface science methods, most of all with STM. This is documented by the large body of work studying only the defects of the bare surface, and the disagreements that remain despite this effort. An additional challenge in studying SAC on MgO is that in addition to the atomic makeup of the surface, the availability of adsorption sites may also depend on the initial charge state of defects, and the final charge state of adatoms may vary accordingly. Furthermore, the experimental limit on film thickness means that some interaction with the underlying metal substrate is hard to rule out entirely. Overall, these challenges make it extremely difficult to compare the thin films to powder catalysts. In our view, MgO is certainly the most challenging of the model systems discussed here, and its interaction with adatoms remain the most poorly understood.

## 6. Copper Oxides (Cu$_2$O)

The Sykes group at Tufts University have been trailblazers in the study of SAC systems using surface science methods. Ever since their seminal paper on "single-atom alloy" systems,[189] they have shown that mechanisms understood using atomically-precise surface science methods can be directly applied to design active powder-based catalysts.[190] Direct collaboration with the Stephanopoulos group, in particular, has shown that this type of synergy can be highly successful. In addition to their work on single atom alloys, Sykes and coworkers have developed a model oxide support based on the monolayer "29" oxide.[191-194] This surface is formed by controlled oxidation of a Cu(111) single crystal in 5×10$^{-6}$ mbar O$_2$ at 650 K, and has a Cu$_2$O stoichiometry. The structure is ring-like, has 29 atoms per unit cell (hence the name) and is close to that of Cu$_2$O(111). It is worth to note that the model is quite complex, and while the agreement between STM and simulated STM is excellent, confirmation by a quantitative structural technique such as LEED-*I*(*V*) or SXRD would provide the ideal basis for the comparison to theory.

Pt adsorbs on the surface as isolated atoms at low temperature, and both XPS and CO-IRAS data suggest that the atoms are close to neutral. Upon heating, isotopically labelled TPD shows that approximately 1/3 of the CO is oxidised to CO$_2$ in a Mars-van Krevelen type process between 300 and 350 K. The remainder of the CO molecules desorb, and irrespective of the reaction pathway the Pt atoms end up under the Cu$_2$O film. This renders them inaccessible for further reactions. Studies of water adsorption on the same system demonstrated dissociative adsorption at the adatom site, and evidence for scrambling between the oxygen atoms from the water and oxide suggest a dynamic rearrangement during TPD acquisition.[193]

The group of Weissenrieder have recently demonstrated a different approach to prepare a stable $CuO_x$-based SAC model system.[195] Evaporating Fe metal directly onto a $Cu_2O(100)$ single crystal surface[196] at room temperature leads to Fe clusters, so they instead deposited $FeCp_2$ molecules, which prevents aggregation of the metal. Based on STM and XPS results, the molecule adsorbs dissociatively into FeCp and Cp fragments. It was then observed that heating the sample to 473 K in a partial pressure of $1\times10^{-6}$ mbar $O_2$ led to the removal of the ligands and an Fe2p signal characteristic of $Fe^{3+}$ cations. STM images clearly show that the surface is covered in isolated protrusions with a uniform height and position within the surface unit cell, and these remain stable up to 573 K. Comparing the data with STM simulations based on DFT calculations, it was concluded that the Fe is coordinated to 2 surface oxygen atoms from the surface, with an additional O ligand provided by the reaction with a gas phase $O_2$ molecule. The O atom is also bound to two surface Cu sites. It should be noted however that the surface structure has not been confirmed by a quantitative structural technique. The authors performed CO oxidation at 473 K, but found that the Fe soon diffuses into the support and becomes inactive. The biggest single takeaway from this paper is that the oxidation of a molecular precursor can lead to stable SAC sites on a surface where metal evaporation does not. It will be very interesting to see if the same method can be applied with other metals and more commonly utilized oxide supports.

## 7. Perspective

This summary of existing data from the surface science community shows quite clearly that metal atoms deposited on low-index metal-oxide surfaces under UHV conditions are typically unstable against agglomeration into clusters. This occurs because the cohesive energy of the metal is higher than the adsorption energy of an adatom on the metal oxide surface. In such cases, isolated atoms can be found at strongly binding defect sites, stabilized by kinetic limitations, or stabilized by an interaction with other surface species (e.g. surface hydroxyl or peroxo groups). There are specific cases in which the metal–host system form a bulk solid solution, as is the case with $Fe_3O_4$, where most metals form a stable $M_xFe_{3-x}O_4$ ferrite. However, if the foreign metal is more oxophilic than the host metal, there will be a tendency to move into the subsurface and ultimately to the bulk to reach a higher coordination to oxygen than can be achieved in the surface layer. Of course, metal atoms in the bulk or even the immediate subsurface are unavailable for reactants, which implies incorporation is likely as much a path to deactivation as thermal sintering. In this context it is important to remember that TEM images do not provide information on the depth of the metal atom being imaged, and that proving a particular atom really resides directly at the surface is extremely difficult.

The lack of virtually any evidence for the occupation of stable bulk-continuation cationic sites is sobering because such sites are often assumed for theoretical modelling of the reaction mechanism. One of the major recommendations to emerge from this analysis is that the barrier for diffusion should be calculated in addition to the adsorption energy before such a geometry is claimed. Then there is the question of environment: The adsorption of reactants can destabilize otherwise stable metal adatoms,[116-118] and there is growing evidence that an interactions with water/OH groups can lead to stabilization.[78,104] It may therefore be necessary to revisit some of the assumptions made in theoretical modelling, testing both more complex adsorption environments and possible co-adsorbates before attempting to predict reaction barriers. Efforts to find true global minima for both surface termination and adatom coordination may be aided by recent advances in applying machine learning methods for surface science.

In powder studies, the SAC systems are typically synthesised using some kind of metal-containing salt, and are further prepared by calcination and/or reduction. In this regard, the work of Wang et al. on $Cu_2O$[195] is particularly interesting because they show quite clearly that the stable $Fe_1$ species they create from ferrocene is an ad-species with additional coordination to oxygen supplied during calcination. Similar processes to stabilize metal atoms have been shown by others,[197-200] and one imagines that such supported geometries with non-native ligands could be commonplace in SAC. In the current authors opinion, it would be prescient to conduct similar studies on common oxide supports to determine to what extent the calcination/reduction step is responsible for adatom stability on surfaces, and to determine to which extent such geometries should be considered in SAC modelling in future.

While not much has been done regarding the mechanism of most reactions using the surface science approach, there is some information regarding CO oxidation. The experience from surface science so far is that the energy required to extract lattice oxygen from low index metal oxide surfaces is high, and too high to account for reactions observed at room temperature and below. On $Fe_3O_4$(001), for example, circa. 550 K is required to remove lattice oxygen from the terrace, so any MvK process likely involves sites with lower-coordinated oxygen such as steps. It is important to note that the MvK mechanism is critically dependent on the oxygen vacancy formation energy, and this depends very much on the surface termination assumed. If one calculates an MvK pathway on an unrealistically oxygen-rich surface termination, it will not be difficult to remove lattice oxygen and the pathway will appear to be energetically favourable. Clearly, theoretical calculations should primarily consider surface terminations that have been shown to exist experimentally. Nonetheless, it may also be wise to more routinely consider SAC environments other than surface continuation sites, which may yield a more oxidized local environment.

It generally seems reasonable to assume a MvK pathway for SAC systems because it is difficult to imagine $O_2$ dissociating at a single atom site, but there is some evidence that associative mechanisms featuring CO-OO intermediates[30] or Eley-Rideal type processes can occur.[201,202] There is also evidence that oxide supported single atoms can dissociate $H_2$,[120] but no report yet of the resulting hydroxyls being used for hydrogenation of any species. It would certainly be advisable for surface science studies to move on from the focus on CO oxidation, particularly as the focus of applied SAC research moves into electrocatalysis.

We have focussed here on oxides that can either be studied as single crystals, or that can be prepared as thin films with a bulk-like structure. Notable exceptions to this are $SiO_2$ and $Al_2O_3$. Like MgO, both are insulating and would provide an interesting contrast to the other systems discussed here in terms of their interaction with adatoms. However, in both cases, the thin films that have been prepared do not appear to be representative of the respective bulk oxides when it comes to adatom adsorption.

For SiO$_2$, quasi-2D monolayer and bilayer films have been grown on a variety of metal substrates.[203,204] Pd adatoms deposited on bilayer SiO$_2$/Ru(0001) at room temperature were shown to diffuse through nanopores in the film to the metal support even in the crystalline phase, in the absence of defects. Au adatoms were slightly more stable, but similarly diffused through the film at defects.[205] On monolayer SiO$_2$/Mo(112), Pd, Ag and Au adatoms have been stabilized by first embedding Pd adatoms in some of the SiO$_2$ nanopores.[206] However, since the adatoms form covalent bonds to the embedded Pd, which in turn is essentially a part of the metal substrate, this is clearly not a good model for single adatoms supported on bulk SiO$_2$. Similar behaviour was observed for Al$_2$O$_3$ thin films outgrown from Ni$_3$Al(111), where single atoms were shown to diffuse to the metal support through pores in the oxide film, but pores already filled by Pd atoms act as cluster nucleation sites.[207] In both cases, it thus seems that the available thin films are not yet suitable as model systems for single atoms supported on oxides, and that studying SAC on these supports would first require developing more bulk-like samples.

One important issue with metal oxide materials that has emerged in the last decade is the presence of polarons, i.e. localized charge associated with lattice distortions.[208] For example, the formation of an oxygen vacancy in a metal oxide leaves two electrons, which will often localize on cation sites, leading to a local distortion of the lattice. Such polarons have been directly imaged in scanning probe experiments,[37] and are known to interact with adsorbates such as CO, substantially changing the adsorption energy.[209] As such, they could also interact with and affect the stability of metal adatoms. Correctly modelling polarons in theoretical calculations is challenging, and almost all of the theoretical work described in this review does not take them into account. This issue will also likely be important for realistic surfaces, because repair of a vacancy by water can result in two electron polarons due to the formation of two surface OH groups. Very recently,[61] the group of Franchini, who are experts in calculating the effect of polarons in metal oxides, have turned their attention to SAC and demonstrated an approach to account for their presence. Specifically, they studied Au, Pt and Rh on TiO$_2$(110), and found that each metal interacts differently with polarons. While polarons are transferred to Au and Pt when they adsorb at $V_O$ sites, Rh adsorbs atop the surface Ti row and polarons are situated in the first subsurface Ti layer.

In recent years there has been a concerted effort to develop the methods of surface science such that they can be applied to UHV-prepared samples moved into ambient pressure and electrochemical regimes. As such, the scene is set for model studies, if suitable model systems can be created. Much of the work with single-atom electrocatalysis is based on carbon or N-doped carbon supports, but as yet there is no surface science study to investigate the coordination effects on stability of such systems, or the adsorption properties. Certainly the ability exists to prepare graphene in situ, and to precisely fabricate N-defects in graphene layers, so depositing metal adatoms and investigating with scanning probe techniques seems like a relatively small but potentially rewarding step. Similarly, single-atom photocatalysis is emerging as an exciting field, and it would be critical to determine what the role of the atom actually is. In principle it can provide new active sites, increase the concentration of existing active sites, provide trap sites that influence recombination rates, and change the electronic structure locally or globally. For this we need to develop model systems based on model photocatalysts. The lack of stability of metal adatoms on titania seems problematic, and we suggest that $Fe_2O_3(1\bar{1}02)$[102-104] (band gap $\approx 2$ eV) can be a fertile model system for fundamental photocatalysis work. Other systems where the surface structures are understood, such as the perovskite $SrTiO_3$, could be ideal if single metal adatoms can be stabilized there.

**Acknowledgements**

GSP and FK acknowledge funding the European Research Council (ERC) under the European Union's Horizon 2020 research and innovation programme (Grant Agreement No. 864628) and by the Austrian Science Fund (FWF, Y847-N20, START Prize).